\documentclass[twocolumn, twocolappendix]{aastex63}

\usepackage[caption=false]{subfig}

\usepackage{amsmath}

\submitjournal{ApJ}

\shorttitle{Highly reddened novae from Palomar Gattini-IR}
\shortauthors{K. De et al.}

\graphicspath{{./}{figures/}}

\begin{document}

\title{A population of heavily reddened, optically missed novae from Palomar Gattini-IR: Constraints on the Galactic nova rate}

\correspondingauthor{Kishalay De}
\email{kde@astro.caltech.edu}

\author[0000-0002-8989-0542]{Kishalay De}
\affil{Cahill Center for Astrophysics, California Institute of Technology, 1200 E. California Blvd. Pasadena, CA 91125, USA.}

\author[0000-0002-5619-4938]{Mansi M. Kasliwal}
\affil{Cahill Center for Astrophysics, California Institute of Technology, 1200 E. California Blvd. Pasadena, CA 91125, USA.}

\author[0000-0001-9315-8437]{Matthew J. Hankins}
\affil{Arkansas Tech University, Russellville, AR 72801, USA}

\author[0000-0003-2835-0304]{Jennifer L. Sokoloski}
\affil{Columbia Astrophysics Laboratory, Columbia University, 550 West 120th Street, New York, NY 10027, USA}

\author[0000-0001-5855-5939]{Scott M. Adams}
\affil{Cahill Center for Astrophysics, California Institute of Technology, 1200 E. California Blvd. Pasadena, CA 91125, USA.}

\author[0000-0003-1412-2028]{Michael C. B. Ashley}
\affil{School of Physics, University of New South Wales, Sydney NSW 2052, Australia}

\author[0000-0001-5491-5423]{Aliya-Nur Babul}
\affil{Department of Astronomy, Columbia University, 550 West 120th Street, New York, NY 10027, U.S.A.}

\author{Ashot Bagdasaryan}
\affil{Cahill Center for Astrophysics, California Institute of Technology, 1200 E. California Blvd. Pasadena, CA 91125, USA.}

\author{Alexandre Delacroix}
\affil{Caltech Optical Observatories, California Institute of Technology, Pasadena, CA 91125, USA}

\author[0000-0002-5884-7867]{Richard Dekany}
\affil{Caltech Optical Observatories, California Institute of Technology, Pasadena, CA 91125, USA}

\author{Timoth\'ee Greffe}
\affil{Caltech Optical Observatories, California Institute of Technology, Pasadena, CA 91125, USA}

\author{David Hale}
\affil{Caltech Optical Observatories, California Institute of Technology, Pasadena, CA 91125, USA}

\author[0000-0001-5754-4007]{Jacob E. Jencson}
\affil{Cahill Center for Astrophysics, California Institute of Technology, 1200 E. California Blvd. Pasadena, CA 91125, USA.}
\affil{Steward Observatory, University of Arizona, 933 North Cherry Avenue, Tucson, AZ85721-0065, USA}

\author[0000-0003-2758-159X]{Viraj R. Karambelkar}
\affil{Cahill Center for Astrophysics, California Institute of Technology, 1200 E. California Blvd. Pasadena, CA 91125, USA.}

\author{Ryan M. Lau}
\affil{Institute of Space \& Astronautical Science, Japan Aerospace Exploration  Agency,  3-1-1 Yoshinodai,  Chuo-ku,  Sagamihara, Kanagawa 252-5210, Japan}

\author[0000-0003-2242-0244]{Ashish Mahabal}
\affil{Cahill Center for Astrophysics, California Institute of Technology, 1200 E. California Blvd. Pasadena, CA 91125, USA.}

\author{Daniel McKenna}
\affil{Caltech Optical Observatories, California Institute of Technology, Pasadena, CA 91125, USA}

\author{Anna M. Moore}
\affil{Research School of Astronomy and Astrophysics, Australian National University, Canberra, ACT 2611, Australia}

\author[0000-0002-6786-8774]{Eran O. Ofek}
\affil{Department of Particle Physics \& Astrophysics, Weizmann Institute of Science, Rehovot 76100, Israel}

\author{Manasi Sharma}
\affil{Department of Physics, Pupin Hall, Columbia University, New York, NY 10027, USA.}

\author[0000-0001-7062-9726]{Roger M. Smith}
\affil{Caltech Optical Observatories, California Institute of Technology, Pasadena, CA 91125, USA}

\author{Jamie Soon}
\affil{Research School of Astronomy and Astrophysics, Australian National University, Canberra, ACT 2611, Australia}

\author[0000-0002-4622-796X]{Roberto Soria}
\affil{National Astronomical Observatories, Chinese Academy of Sciences, Beijing 100012, China}

\author{Gokul Srinivasaragavan}
\affil{Cahill Center for Astrophysics, California Institute of Technology, 1200 E. California Blvd. Pasadena, CA 91125, USA.}

\author[0000-0002-1481-4676]{Samaporn Tinyanont}
\affil{University of California Santa Cruz, 1156 High St, Santa Cruz, CA 95064}

\author[0000-0001-9304-6718]{Tony Travouillon}
\affil{Research School of Astronomy and Astrophysics, Australian National University, Canberra, ACT 2611, Australia}

\author[0000-0003-0484-3331]{Anastasios Tzanidakis} 
\affil{Cahill Center for Astrophysics, California Institute of Technology, 1200 E. California Blvd. Pasadena, CA 91125, USA.}

\author[0000-0001-6747-8509]{Yuhan Yao}
\affil{Cahill Center for Astrophysics, California Institute of Technology, 1200 E. California Blvd. Pasadena, CA 91125, USA.}

\begin{abstract}
The nova rate in the Milky Way remains largely uncertain, despite its vital importance in constraining models of Galactic chemical evolution as well as understanding progenitor channels for Type Ia supernovae. The rate has been previously estimated in the range of $\approx10-300$\,yr$^{-1}$, either based on extrapolations from a handful of very bright optical novae or the nova rates in nearby galaxies; both methods are subject to debatable assumptions. The total discovery rate of optical novae remains much smaller ($\approx5-10$\,yr$^{-1}$) than these estimates, even with the advent of all-sky optical time domain surveys. Here, we present a systematic sample of 12 spectroscopically confirmed Galactic novae detected in the first 17 months of Palomar Gattini-IR (PGIR), a wide-field near-infrared time domain survey. Operating in $J$-band ($\approx1.2$\,$\mu$m) that is relatively immune to dust extinction, the extinction distribution of the PGIR sample is highly skewed to large extinction values ($> 50$\% of events obscured by $A_V\gtrsim5$\,mag). Using recent estimates for the distribution of mass and dust in the Galaxy, we show that the observed extinction distribution of the PGIR sample is commensurate with that expected from dust models. The PGIR extinction distribution is inconsistent with that reported in previous optical searches (null hypothesis probability $< 0.01$\%), suggesting that a large population of highly obscured novae have been systematically missed in previous optical searches. We perform the first quantitative simulation of a $3\pi$ time domain survey to estimate the Galactic nova rate using PGIR, and derive a rate of $\approx 46.0^{+12.5}_{-12.4}$\,yr$^{-1}$. Our results suggest that all-sky near-infrared time-domain surveys are well poised to uncover the Galactic nova population.
\end{abstract}


\section{Introduction} \label{sec:intro}

A nova outburst arises from a thermonuclear runaway on the surface of a white dwarf caused by unstable nuclear burning in a degenerate layer accreted from a companion \citep{Bode2008,Starrfield2016, DellaValle2020, Chomiuk2020}. As factories for the nucleosynthesis of elements as well as crucial phases in the evolution of binary low mass stars, the total rate and demographics of novae in the Milky Way are important to constrain the chemical evolution of the Galaxy. Nucleosynthesis in novae (see \citealt{Gehrz1998} and \citealt{Jose2006} for reviews) plays a crucial role in the synthesis of isotopes like $^7$Li, $^{22}$Na, $^{26}$Al and $^{15}$N \citep{Romano2003, Prantzos2012}. Novae have also been long suggested as possible progenitors of Type Ia supernovae \citep{Shafter2015, Soraisam2015, Starrfield2020a, Starrfield2020b}, especially with the discovery of rapidly recurrent novae such as M31N 2008-12a \citep{Tang2014, Darnley2014, Henze2015a, Henze2015b}. Such recurrent novae likely contain very massive white dwarfs imminent for a complete thermonuclear supernova in $\lesssim 10^6$\,years \citep{Kato2014, Hillman2016}, thus providing unique windows into the still poorly understood progenitors of Type Ia supernovae (see \citealt{Maoz2014, Darnley2020} for a review).

Despite its importance, the nova rate in the Galaxy remains observationally poorly constrained. Previous estimates for the Galactic nova rate primarily use two classes of techniques. One class uses the observed extragalactic nova rate from nearby galaxies to scale to the Milky Way's estimated K-band luminosity, resulting in rates over a range of $\approx 10 - 50$\,yr$^{-1}$ \citep{Ciardullo1990, vanDenBergh1991, DellaValle1992b, DellaValle1994, Shafter2000, Darnley2006}. Unlike direct rate measurements of Galactic novae, the extragalactic estimates suffer from uncertainties in the differences between the star formation history, structure and stellar population of the  Milky Way and external galaxies \citep{DellaValle1994, DellaValle1994b, Shafter2002}.

The second class of techniques uses the statistics of a handful of very bright (naked-eye and nearby) Galactic novae to extrapolate the rate to the entire Galaxy, resulting in estimates in a large range of $\approx 30 - 300$\,yr$^{-1}$ \citep{Allen1954, Sharov1972, Liller1987, Hatano1997, Shafter1997, Shafter2002, Shafter2017, Ozdonmez2018}. However, all previous estimates are subject to the poorly quantified selection effects of the discovery of even the brightest optical novae (e.g. \citealt{Schaefer2014, Shafter2017}). More recently, \citet{Mroz2015} presented a list of likely nova candidates\footnote{Most candidates in their sample were not confirmed with spectroscopy, and thus have contamination from other types of large amplitude variables like dwarf novae.} in the Galactic bulge from the Optical Gravitational Microlensing Experiment (OGLE; \citealt{Udalski1992}) survey, and used it to estimate a bulge rate of $13.8 \pm 2.6$\,yr$^{-1}$. 

Based on models of the distribution of mass and dust in the Galaxy, \citet{Shafter2017} suggest that the the number of detectable novae in optical surveys (to a depth of $\approx 17$\,mag) should be $50^{+31}_{-23}$\,yr$^{-1}$ depending on the assumed completeness for the brightest novae ($m <2$; see their Figure 8 and see also \citealt{Hatano1997}). Despite the emergence of wide-field optical surveys that can routinely survey the entire sky to this depth, the discovery rate of novae\footnote{\url{https://asd.gsfc.nasa.gov/Koji.Mukai/novae/novae.html}}$^,$\footnote{\url{https://github.com/Bill-Gray/galnovae/blob/master/galnovae.txt}} has remained much smaller at $\approx 5 - 10$\,yr$^{-1}$. Thus, either the nova rate in the Milky Way has been grossly overestimated, or a large fraction of novae are missed or misidentified in optical searches (e.g. \citealt{Hounsell2010}). Alternatively, since the optical rate estimates are critically subject to uncertainties regarding the distribution of obscuring dust, many novae could be highly reddened and undetectable in optical searches (e.g. \citealt{Hounsell2011}). In particular, we note that recent estimates of the Galactic dust distribution \citep{Green2019} reveal rich structures that are not captured by the simple double-exponential models used in previous works.

Given the extreme dust obscuration in the optical bands \citep{Cardelli1989}, the lower effects of extinction in the near-infrared (NIR\footnote{For the rest of this work, we refer to wavelengths $1 - 3\mu$m as NIR bands}) bands make them ideally suited to search for these eruptions. However, large area surveys in the NIR bands have been prohibitively expensive due to the bright sky foreground as well as the high cost of detectors. The Vista Variables in the Via Lactea (VVV; \citealt{Catelan2011}) survey was one of the largest such experiment carried out previously, involving a deep (to $K\approx 18$\,mag) and slow ($\approx 10 - 30$\,epochs per year) time domain survey of a fraction of the southern Galactic bulge and disk. In particular, they reported $\approx 20$ dust obscured nova candidates (e.g. \citealt{Saito2012, Saito2013a, Saito2013b, ContrerasPena2017a, ContrerasPena2017b}) from their search. However, these candidates were not confirmed with real-time spectroscopic follow-up due to their discovery in archival images, and likely contain contamination from large amplitude young stellar object outbursts and foreground dwarf novae \citep{ContrerasPena2017a}.

In this paper, we present a sample of 12 spectroscopically confirmed novae detected in the first 17 months of the Palomar Gattini-IR (PGIR) NIR time domain survey \citep{Moore2019, De2020a}. PGIR is a robotic, wide-field time domain survey at Palomar observatory using a 25\,sq.\,deg. J-band camera to survey the entire northern visible sky ($\delta > -28.9^\circ$; $\approx 15000$\,sq.\,deg.) at a cadence of $\approx 2$ nights. We use this sample together with detailed simulations of the PGIR survey to construct the first constraints on the Galactic nova rate using a NIR discovery engine. In Section \ref{sec:candidates}, we describe the techniques for identification of large amplitude transients in the PGIR transient stream and the sample of identified novae. In Section \ref{sec:extinction}, we compare the extinction distribution of the PGIR nova sample to that in previous optical samples to highlight a population of highly obscured novae that have been systematically missed in optical searches. In Section \ref{sec:rate}, we present detailed simulations of the PGIR survey and detection efficiency of the PGIR pipeline to present constraints on the Galactic nova rate. In Section \ref{sec:discussion}, we discuss the assumed parameters for the specific nova rates, luminosity function and luminosity-width relationships in the context of variations in the derived nova rate. We conclude with a summary of our findings in Section \ref{sec:summary}.

\section{Candidate selection}
\label{sec:candidates}

The median $5\sigma$ sensitivity of PGIR is $14.8$\,Vega\,mag ($\approx 15.7$\,AB\,mag; \citealt{De2020a}) outside the Galactic plane. The sensitivity is limited by confusion to $\approx 1-2$ mag shallower in the Galactic plane due to the large ($\approx 8.7$\,\arcsec) pixel scale of the detector ($2048 \times 2048$ pixels). The typical saturation magnitude for the detector was $\approx 8.5$ Vega mag until May 2020; a modification of the readout electronics improved the dynamic range to $\approx 6.0$ Vega mag at the bright end for later data \citep{De2020c}. A dedicated data processing system produces science quality stacked images together with transient candidates identified from subtractions against template images (using the ZOGY algorithm; \citealt{Zackay2016}) in real-time. Following a deep-learning based machine-learning classification system, the transient candidates are vetted by human scanners on a daily basis for photometric and spectroscopic follow-up. 

\subsection{Identification of nova candidates}

This work considers survey data acquired in the first 17 months of the survey -- between 2 July 2019 (the start date of the survey when the reference image construction was completed) and 30 Nov 2020 (the end of the 2020 Galactic bulge season in the northern hemisphere). We carried out a systematic search for large amplitude transients in the PGIR stream to search for Galactic novae. Candidate transients were identified as sources with at least three positive detections (i.e., the source flux has increased from the reference image) satisfying the following criteria:
\begin{enumerate}
    \item Detected with signal-to-noise ratio (SNR) $> 8$ without saturation. We require that the detection epochs should be separated by $> 1$\,day to eliminate contamination from solar system objects.
    \item A real-bogus classification score $RB > 0.5$, that has been tested to produce a false positive rate of $1.6$\% and a false negative rate of $1.4$\% \citep{De2020a}.
    \item The transient is either hostless (i.e., no known 2MASS counterpart within a radius of 10\,arcsec) or has a large amplitude (i.e., is at least 3 mags brighter than any 2MASS counterparts within a radius of 10\,arcsec). The choice of the amplitude and radius were defined to exclude the large contamination from variable stars in the Galactic plane given the coarse pixel scale ($\approx 8.7$\,arcsec) of the PGIR detector. 
\end{enumerate}

These selection criteria result in an average of $\approx 50-100$ candidates per night to be examined with human vetting. The majority of false positives arise from astrometric residuals on bright stars in parts of the detector where the point spread function is elongated and sub-optimal (see discussion of image quality variation in \citealt{De2020a}).

\begin{table*}[]
    \centering
    \footnotesize
    \begin{tabular}{lcccccccccc}
    \hline
        PGIR Name & Variable & RA & Dec & MJD & Peak J & Phot-class & Spec-class & $A_{V, c}$ & $A_{V, s}$ & $A_{V, t}$\\
        & & J2000 & J2000 & First Det. & Vega mag & & & mag & mag & mag \\
        \hline
        PGIR\,20ekz & V3731\,Oph & 17:38:35.1 & -25:19:02.9 & $58685.25$ & $8.71 \pm 0.01$ & S-class & He/N? & $6.4$ & -- & $5.5$\\
        PGIR\,19bte & V2860\,Ori & 6:09:57.4 & 12:12:24.8 & $58735.52$ & $12.07 \pm 0.03$ & D-class & He/N  & $1.0$ & $1.7$ (K $7699$) & $2.1$\\
        PGIR\,19bgv & V569\,Vul & 19:52:08.2 & 27:42:21.1 & $58716.28$ & $8.42 \pm 0.01$ & S-class & He/N  & $8.9$ & $10.0$ (K $7699$) & $9.6$\\
        PGIR\,19brv & V2891\,Cyg & 21:09:25.5 & 48:10:51.9 & $58743.25$ & $8.83 \pm 0.01$ & F-class & Fe-II  & $7.3$ & $12.2$ (K $7699$) & $12.1$ \\
        PGIR\,19fai & V3890\,Sgr & 18:30:43.1 & -24:01:10.5 & $58744.14$ & $9.92 \pm 0.01$ & -- & Symbiotic & -- & $1.9$ (Na D1) & $1.5$\\
        PGIR\,20dcl & V659\,Sct & 18:39:59.7 & -10:25:43.1 & $58904.56$ & $7.59 \pm 0.01$ & J-class & Hybrid & $2.9$ & $4.0$ (K $7699$) & $5.7$\\
        PGIR\,20duo & V2000\,Aql & 18:43:53.4 & 0:03:51.7 & $58981.34$ & $10.38 \pm 0.01$ & S-class & Fe-II & $9.4$ & -- & $8.4$\\
        PGIR\,20dsv & V6567\,Sgr & 18:22:45.2 & -19:36:02.6 & $59001.43$ & $9.01 \pm 0.01$ & O-class & Fe-II & $4.6$ & $4.3$ (DIB $5780$) & $4.7$ \\
        PGIR\,20eig & V2029\,Aql & 19:14:27.0 & 14:44:32.0 & $59043.26$ & $10.41 \pm 0.02$ & C-class & Fe-II & $6.1$ & $5.7$ (DIB $5780$) & $5.9$ \\
        PGIR\,20emj & V1391\,Cas & 0:11:42.8 & 66:11:19.1 & $59076.48$ & $8.04 \pm 0.01$ & D-class & Fe-II & $4.7$ & $4.6$ (DIB $5780$) & $4.0$\\
        PGIR\,20evr & V6593\,Sgr & 17:54:59.9 & -21:22:41.3 & $59124.11$ & $6.98 \pm 0.01$ & -- & Fe-II & $4.2$ & -- & $4.7$  \\
        PGIR\,20fbf & V1112\,Per & 04:29:18.7 & +43:54:21.5 & $59180.22$ & $6.22 \pm 0.01$ & D-class & R-Hybrid & $2.3$ & $2.4$ (K $7699$) & $2.1$ \\
        \hline
    \end{tabular}
    \caption{Summary of novae detected in the first 17 months of the PGIR survey (see Appendix \ref{sec:app_nova} for details). The column MJD Detection denotes the first detection of the nova in PGIR difference images, while the Peak J Mag denotes the brightest J magnitude in the nova light curve (note that the PGIR survey may have missed the peak light curve for some novae). The Phot-Class and Spec-Class columns denote the photometric and spectroscopic class of the nova based on the classification schemes of \citet{Strope2010} and \citet{Williams1992} respectively. We caution that the light curve parameters and classifications are limited by the photometric coverage available for each nova. The $A_{V, c}$ and $A_{V, s}$ columns indicate the estimated extinction towards the nova using the photometric color and spectroscopic methods respectively, while the $A_{V,t}$ column indicates the total Galactic extinction along this direction estimated in \citet{Schlafly2011}. For each spectroscopic reddening measurement, we indicate the absorption feature used to estimate the reddening. We caution that the high extinction inferred towards most events lie in a regime beyond the well calibrated range for these features.}
    \label{tab:nova_list}
\end{table*}

 \begin{figure*}[!ht]
    \centering
    \includegraphics[width=0.8\textwidth]{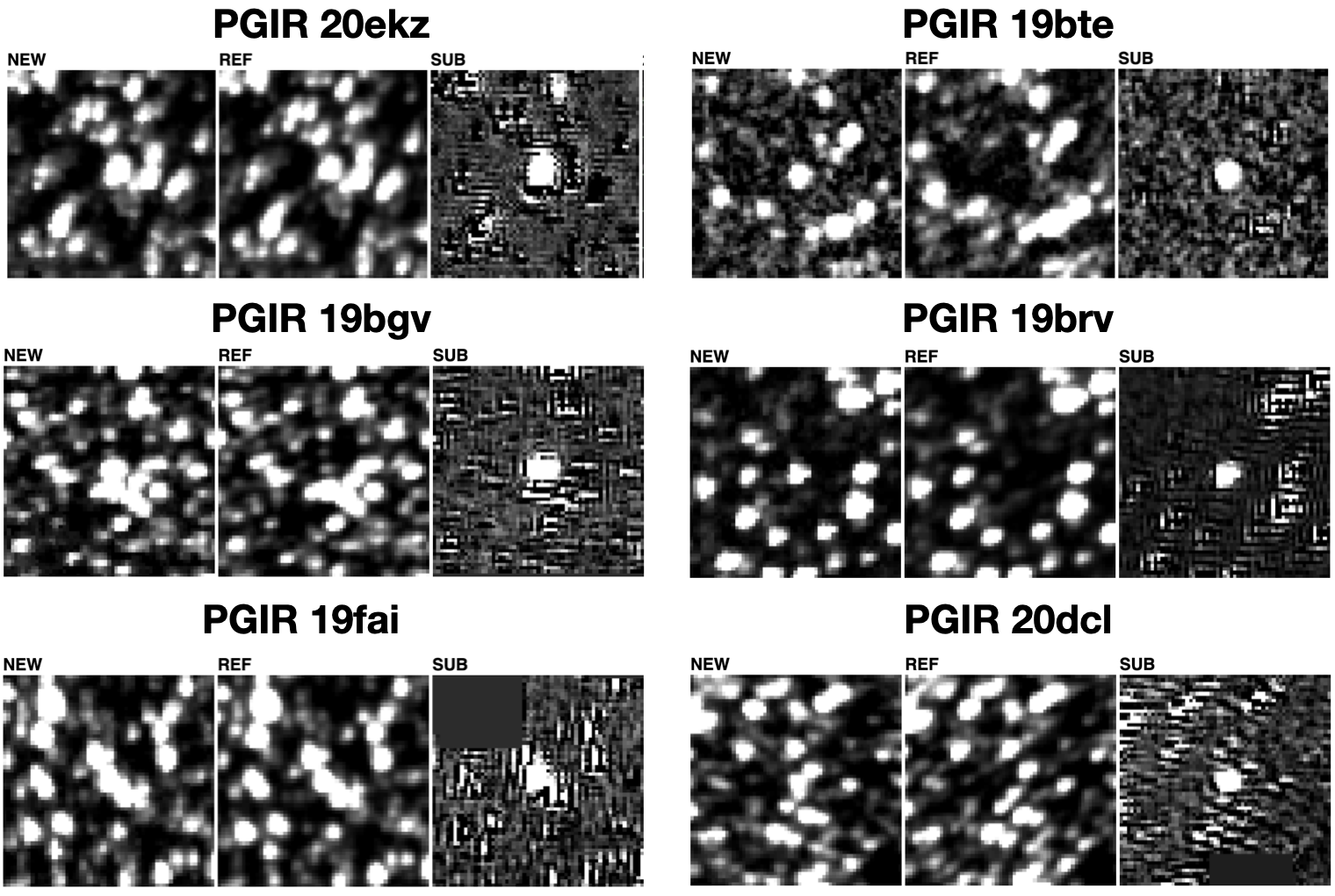}
    \includegraphics[width=0.8\textwidth]{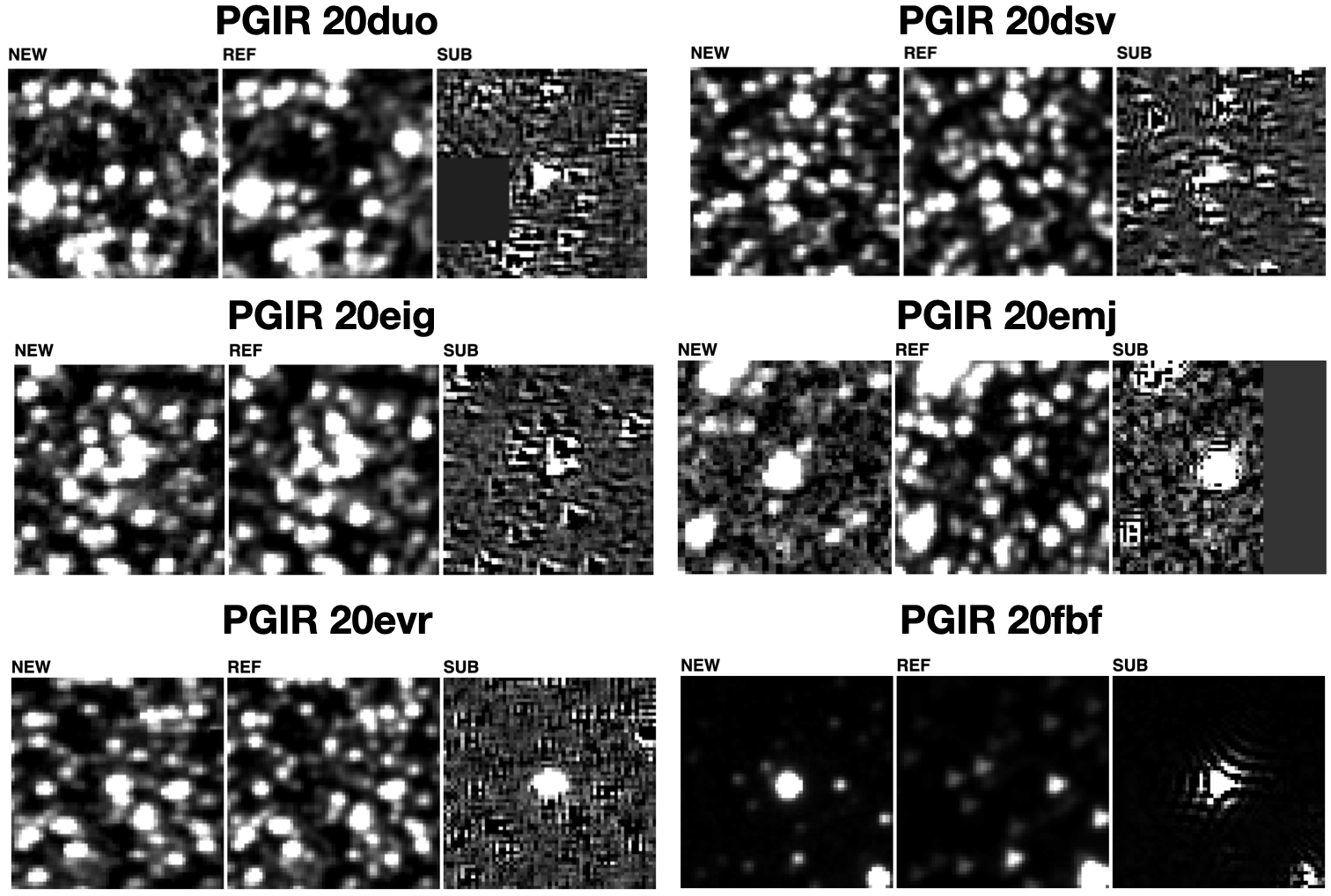}
    \caption{Cutout triplets of science (new), reference (ref) and difference (sub) images of novae detected in the the first 17 months of the PGIR survey. The name of the nova is indicated on top of each triplet. The novae are clearly detected as bright transients at the center of the subtracted image in each panel. The variation in the image quality for different sources is due to the variation of the optical point spread function across the detector plane \citep{De2020a}.}
    \label{fig:cutouts}
\end{figure*}

\subsection{Confirmation and Follow-up}

All transients that pass these criteria were cross-matched against external catalogs for prior known classifications. In addition, we identified large amplitude regular/semi-regular variables using independent criteria based on known variable counterparts in archival Wide-field Infrared Survey Explorer (WISE; \citealt{Mainzer2011, Mainzer2014}) images. These sources will be presented as separate publications focusing on R Coronae Borealis variables \citep{Karambelkar2020} and Young Stellar Objects (\citealt{Hankins2020}, Hankins et al. in prep.). For sources determined to be bonafide eruptive transients, we assigned photometry and spectroscopy using the Spectral Energy Distribution Machine (SEDM; \citealt{Blagorodnova2018}) spectrograph mounted on the Palomar 60-inch telescope for rapid confirmation and characterization. The SEDM data were reduced using the automated \texttt{pysedm} pipeline \citep{Rigault2019}. In cases where the source was heavily reddened or in very crowded fields preventing a reliable spectrum from SEDM, we used optical / NIR spectroscopy on larger telescopes for classification.

A total of 44 large amplitude transients passed our selection criteria and human vetting. 11 sources were determined to be active galactic nuclei known from prior surveys, 3 were classified as bright extragalactic supernovae (e.g. \citealt{DeA2020f}; to be presented in Srinivasaragavan et al. in prep.), 7 sources were determined to be outbursts of young stellar objects (e.g. \citealt{Hankins2020}, presented in \citealt{Hillenbrand2021} and Hankins et al. in prep.) and 1 source was a previously known low mass X-ray binary in outburst \citep{Hankins2019}. We identified 2 microlensing events using their characteristic photometric evolution (e.g. \citealt{DeA2019b}; to be presented in Mroz et al. in prep.)

\begin{figure*}[!ht]
    \centering
    \includegraphics[width=0.97\textwidth]{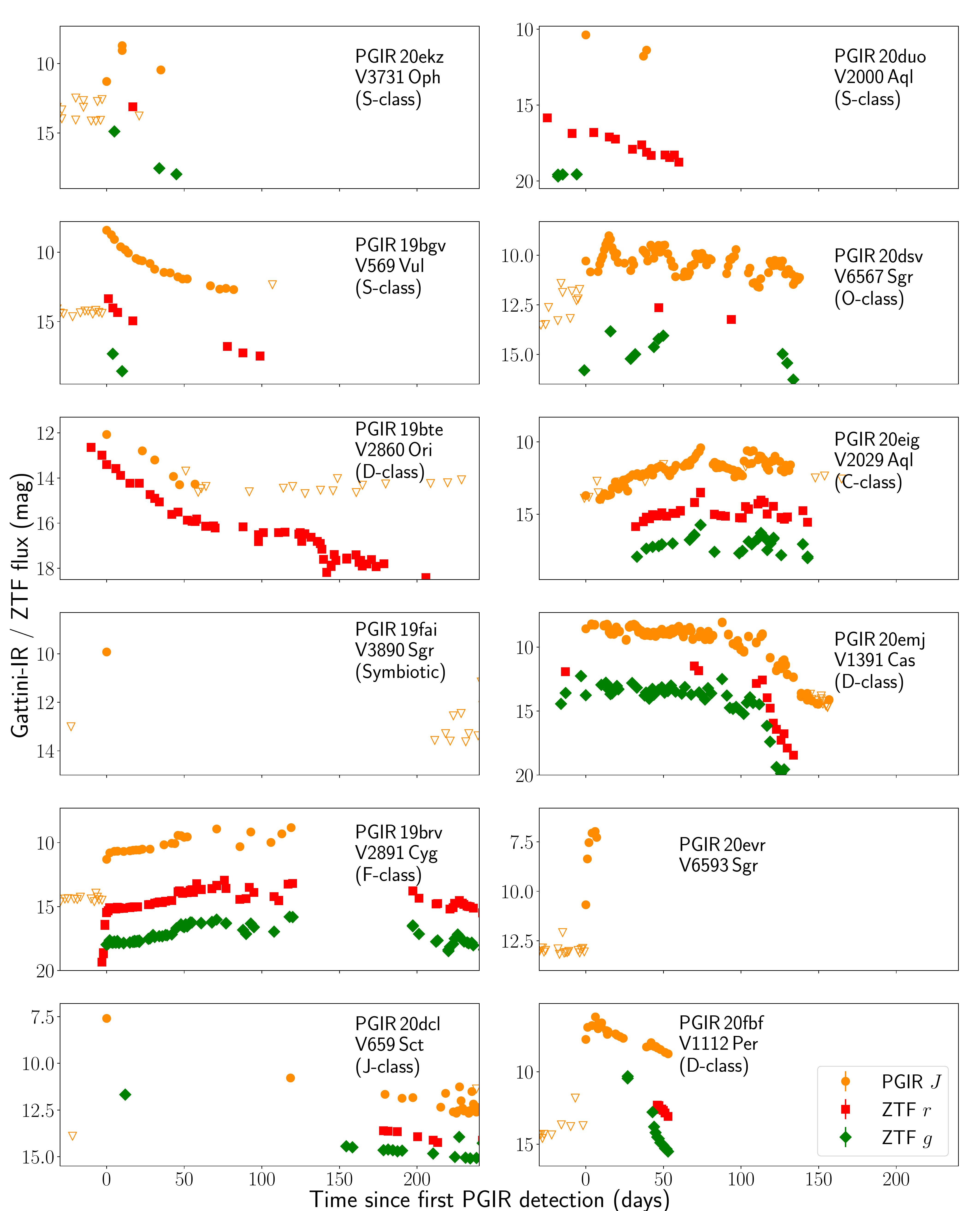}
    \caption{Multi-color light curves of the PGIR nova sample from Gattini-IR ($J$-band as orange circles) and ZTF ($g$-band as green diamonds and $r$-band as red squares). Hollow inverted triangles show $5\sigma$ upper limits from PGIR. In each panel, the name of the nova is shown along with its photometric class based on the classification scheme in \citet{Strope2010}. The lack of post-peak photometric data precludes a classification for PGIR\,20evr.}
    \label{fig:photometry}
\end{figure*}

In particular, dwarf novae \citep{Warner1995} from cataclysmic variables bear striking photometric similarities to nova outbursts and represent a contaminant in this search. Despite being substantially more abundant than novae, their relatively low luminosity ($M \approx 3 - 5$ at peak) distinguishes them from nova outbursts. We thus reject transients that have peak apparent magnitude fainter than that expected for the lowest luminosity novae ($M \approx -4$ at peak) at the farthest edge of the Galaxy (distance modulus of $\approx 16.5$) after accounting for the integrated $J$-band extinction along the line of sight \citep{Schlafly2011}. We further used archival classifications from SIMBAD as well as data from other time domain surveys (e.g., prior eruptions detected in the optical) to reject these objects\footnote{In order to ensure that recurrent novae were not mistaken as repeating dwarf nova outbursts, we confirmed the classification of all sources that were bright enough to be novae at the farthest edge of the Galaxy by examining publicly available reports and spectra}. Together, we identified a total of 9 dwarf nova outbursts in our search. Appendix \ref{sec:app_nova} summarizes two previously unknown dwarf novae that we followed up and confirmed with spectroscopy.

In this paper, we focus on the sample of 12 spectroscopically confirmed novae found from this search. Eleven out of the 12 novae discussed here were selected using the criteria discussed above; the only exception is the 2019 eruption of the symbiotic-like recurrent nova V3890\,Sgr \citep{Schaefer2010, Page2020}. The eruption of V3890\,Sgr was detected in its early stages in the PGIR data but did not pass our selection criteria due to saturation of the detector near peak brightness, and we include it here for completeness. In cases where the nova was first identified in PGIR survey data, the detection and spectroscopic confirmation were immediately announced to the community via the Transient Name Server\footnote{\url{https://wis-tns.org}} and The Astronomer's Telegram \citep{DeA2019, DeA2020a, DeA2020b, DeA2020c, DeA2020d}.

Detection images of identified PGIR novae are shown in Figure \ref{fig:cutouts} and their light curves shown in Figure \ref{fig:photometry}. Table \ref{tab:nova_list} summarizes the properties of novae discussed in this paper, while Appendix \ref{sec:app_nova} presents a brief summary of the initial identification and properties of each nova. All the novae presented here were identified and followed up in real-time during the eruption, with the exception of V3731\,Oph, which was confirmed with a late-time spectrum in September 2019 from an archival search of early PGIR survey data \citep{DeA2020e}. Our search detected all but one nova reported publicly in the PGIR observing footprint during the survey period considered. The confirmed nova V670 Ser \citep{AydiA2020b, Taguchi2020}  was not detected due to its eruption shortly after solar conjunction when PGIR was not observing the field.

We accumulated multi-color photometry of each nova by performing forced photometry on the PGIR difference images, and accumulated publicly available $r$-band and $g$-band data from the Zwicky Transient Facility optical time domain survey \citep{Bellm2019, Masci2019}. Figure \ref{fig:photometry} shows a collage of the optical/NIR light curves of the novae presented here. We classify the light curves using the combined optical and NIR dataset based on the scheme presented in \citet{Strope2010} and discuss features of individual objects in Appendix \ref{sec:app_nova}. As shown in Figure \ref{fig:photometry}, the PGIR nova sample consists of diverse photometric classes encompassing all the types discussed in \citet{Strope2010}.

\begin{figure*}[!ht]
    \centering
    \includegraphics[width=0.92\textwidth]{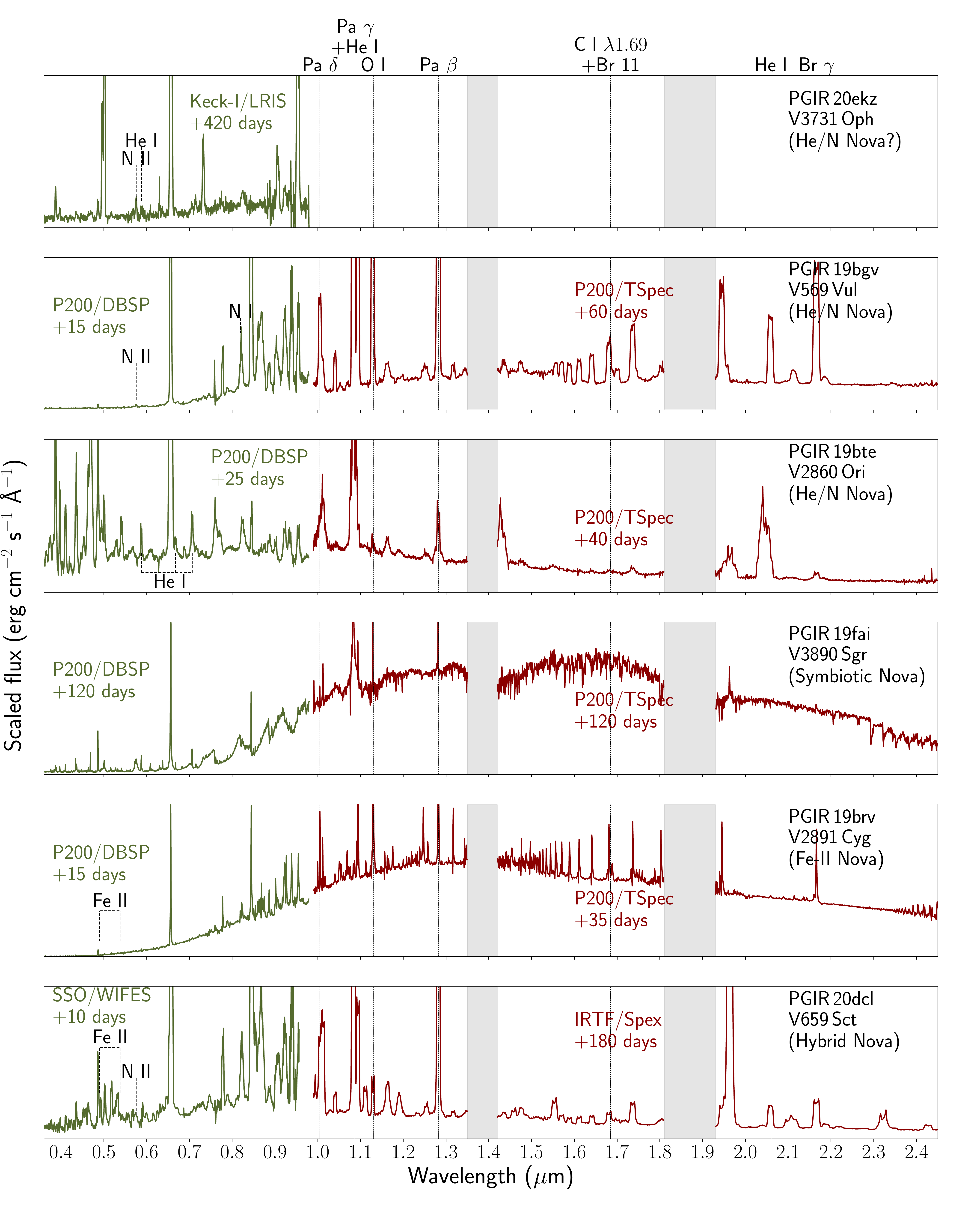}
    \caption{Medium resolution optical and near-infrared spectra of the sample of novae discussed in this work. Bands of low atmospheric transmission in the NIR are blocked out as shaded bands. In each panel, we indicate the instrument used and the phase of the spectrum with respect to the date of first detection in Table \ref{tab:nova_list}. We show the name(s) of the nova along with their spectroscopic classification according to the scheme of \citet{Williams1992}. In each panel, we also highlight the primary spectroscopic features used to identify the nova type (R. Williams, priv. comm.), with the exception of V3890\,Sgr which is a known recurrent symbiotic-like recurrent nova \citep{Schaefer2010}. See Appendix \ref{sec:app_nova} for more details. }
    \label{fig:spec_1}
\end{figure*}

\begin{figure*}[!ht]
    \ContinuedFloat
    \centering
    \includegraphics[width=0.92\textwidth]{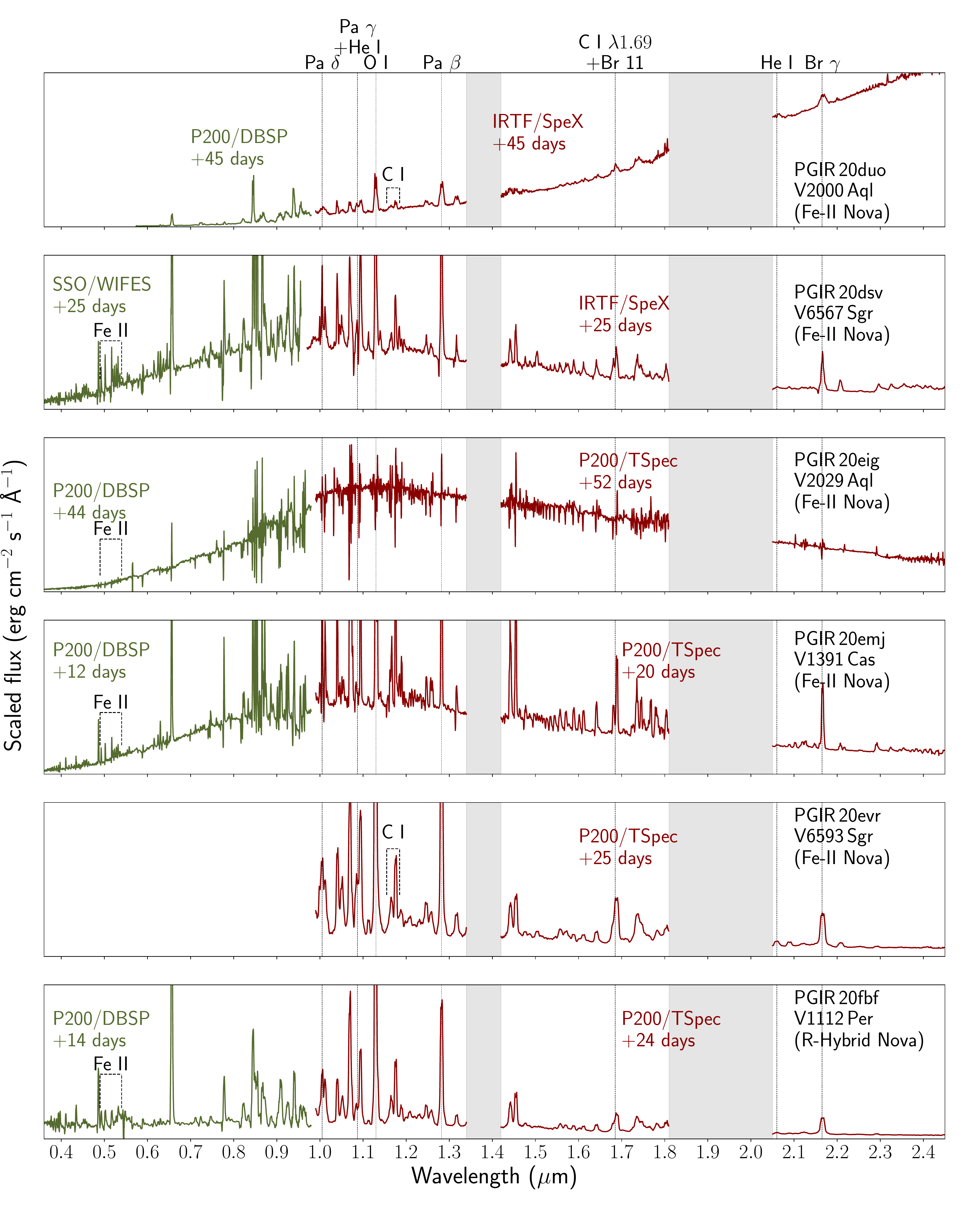}
    \caption{Continued}
\end{figure*}

In order to determine spectroscopic classifications of confirmed novae \citep{Williams1992}, we obtained medium resolution optical and infrared spectroscopic follow-up (see Figure \ref{fig:spec_1}) using the Palomar 200-inch telescope (P200) at Palomar observatory, the NASA Infrared Telescope Facility (IRTF) and Keck-I telescopes at Mauna Kea, and the 2.3 m telescope at Siding Spring Observatory (SSO). The P200 data were acquired using the optical Double Spectrograph (DBSP; \citealt{Oke1982}) and near-infrared Triple spectrograph (TSpec; \citealt{Herter2008}). The DBSP data were reduced using the \texttt{pyraf-dbsp} pipeline \citep{Bellm2016b} while the TSpec data were reduced using the \texttt{spextool} \citep{Cushing2004} and \texttt{xtellcor} \citep{Vacca2003} packages. The IRTF data were acquired with the SpeX instrument \citep{Rayner2003} in the SXD mode ($\approx 0.7 - 2.5\mu$m) as part of programs 2020A111 and 2020B087 (PI: K. De). The data were reduced using the \texttt{spextool} \citep{Cushing2004} and \texttt{xtellcor} \citep{Vacca2003} packages. The Keck spectrum of PGIR\,20ekz was acquired using the Low Resolution Imaging Spectrometer (LRIS; \citealt{Oke1995}) and was reduced using the \texttt{lpipe} package \citep{Perley2019}. The SSO data were acquired using the Wide-field Spectrograph (WiFeS; \citealt{Dopita2010}) on the ANU 2.3 m telescope and reduced using the PyWiFeS pipeline \citep{Childress2014}.

\section{The extinction distribution}
\label{sec:extinction}

\begin{figure*}[!htp]
    \centering
    \includegraphics[width=\textwidth]{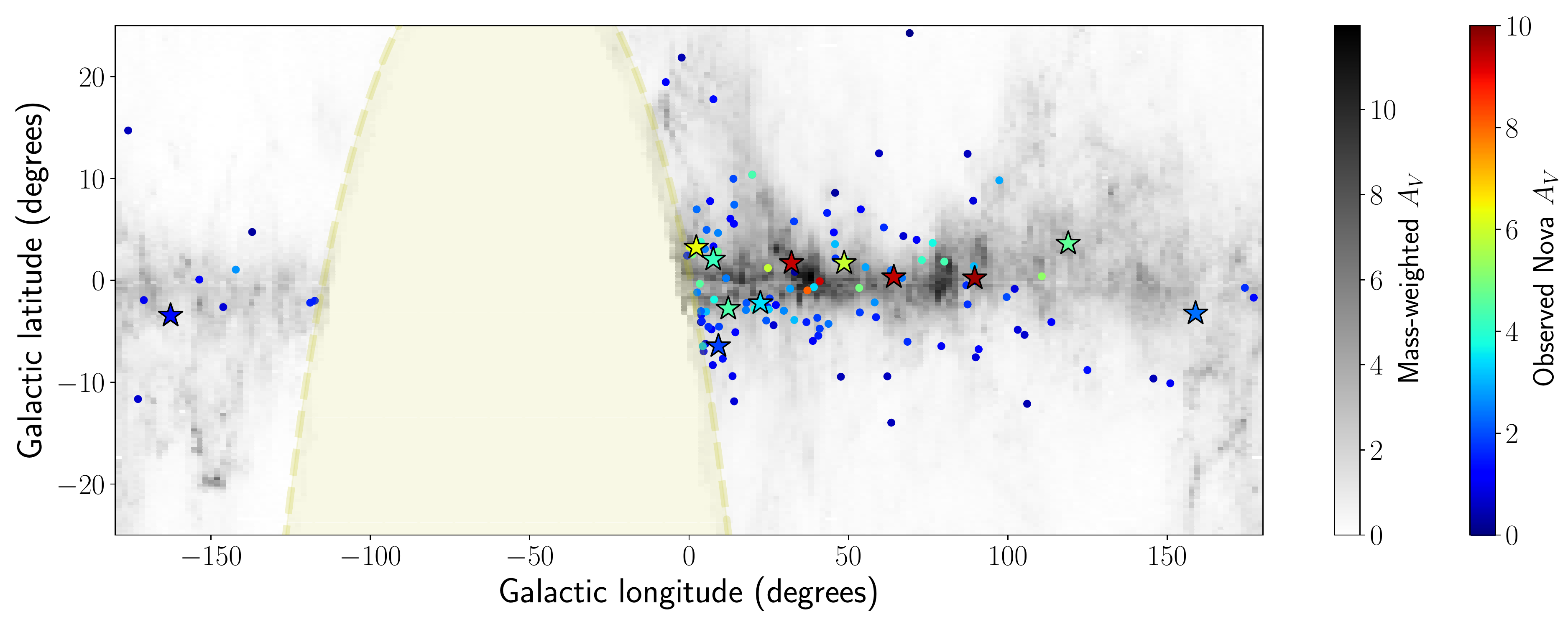}
    \caption{Sky distribution of the PGIR nova sample (in galactic coordinates) shown as stars together with previous optical nova samples (in circles) from \citealt{Ozdonmez2016} and \citealt{Ozdonmez2018}. In the background, we overlay a grey-shaded map of the mass-weighted dust distribution (see text) in the Galaxy using the Milky Way mass model from \citet{Cautun2020} and dust distribution from \citet{Green2019}. For each nova, we also show the inferred extinction towards the nova indicated by the rainbow color bar on the right. For the PGIR novae, we use the average of the extinction estimated from photometric and spectroscopic methods. The yellow shaded region lies south of the $\delta = -28.9^\circ$ viewing limit of PGIR, and is not included in the \citet{Green2019} maps (which are based on PS1 images).}
    \label{fig:galdist}
\end{figure*}

A striking feature of the PGIR novae is the abundance of highly reddened novae (as is evident from both photometry and spectroscopy; Figures \ref{fig:photometry}, \ref{fig:spec_1}). We thus compare this sample to previous optically selected nova samples by deriving extinction estimates for each object using both photometric and spectroscopic methods. 

\subsection{Extinction from photometric color evolution}

As in several previous works (e.g. \citealt{Hachisu2014, Hachisu2016}), we use the photometric color evolution of novae to estimate the line-of-sight extinction. While there is some uncertainty on the exact template phase as well as color during the eruption, the intrinsic color of novae has been estimated to be $B - V \approx 0.0 - 0.2$\,mag \citep{vandenBergh1987, Miroshnichenko1987, Hachisu2014} and consistent with the colors of A5V stars \citep{Shafter2009}. For objects in our sample with multi-color photometry, we use the intrinsic colors of A5V stars from \citet{Kraus2007} of $(g - J)_0 = 0.35$\,mag and $(r - J)_0 = 0.42$\,mag to estimate extinctions using a \citet{Cardelli1989}\footnote{Throughout this paper, we assume $R_V = A_V / E(B-V) = 3.1$ to relate the extinction across the optical and NIR bands.} extinction law as
\begin{equation}
\centering
\begin{split}
    A_V = 1.06\, E(g - J) = 1.75 \, E(r - J)\\
    = 1.06\, ((g-J) - (g - J)_0)
    = 1.06 \, ((g-J) - 0.35)\\
    = 1.75 \, ((r-J) - (r - J)_0) = 1.75 \, ((r-J) - 0.42)
    \end{split}
\end{equation}
where $(g-J)$ and $(r-J)$ are the observed nova colors near peak. We expect an uncertainty of $\approx 0.2$\,mag using this method, based on the spread in estimated values of the intrinsic $B-V$ color. The extinctions derived using this method are given in Table \ref{tab:nova_list} together with the total integrated extinction expected along this line of sight from the maps of \citet{Schlafly2011}. The estimates derived using this method are generally consistent with spectroscopic techniques in cases where the spectroscopic features are not saturated.

\subsection{Extinction from spectroscopic features}

As in previous works (e.g. \citealt{Ozdonmez2016, Ozdonmez2018}), we use the equivalent widths of the Na I D ($\lambda\lambda5890.0,5895.9$\,\AA) and K I ($\lambda 7699$\,\AA) lines to estimate the line-of-sight extinction to each nova using the calibration in \citet{Munari1997}. We also search for absorption from the diffuse interstellar band (DIB) features to constrain the extinction using previously established relations \citep{Yuan2012}. In several cases, the novae exhibit broad / P-Cygni features in the Na and K lines precluding measurements of the interstellar features in those objects given our relatively low resolution spectra and hence we attempt to use the DIB features. We note that the K I line has been shown to be more sensitive at higher extinctions unlike the Na I lines which saturate beyond $E(B-V) \gtrsim 1.5$\,mag \citep{Munari1997}; hence, we use the K I lines where detected. Regardless, the highly reddened nature of most of these sources ($E(B-V) \gtrsim 1.5$\,mag) require extrapolation of the nominal relationships beyond their previously established range. The derived extinctions are given in Table \ref{tab:nova_list} and a description of the derived extinction values are given in Appendix \ref{sec:app_nova}. 

\subsection{Comparison to optically discovered novae}

\begin{figure}
    \centering
    \includegraphics[width=\columnwidth]{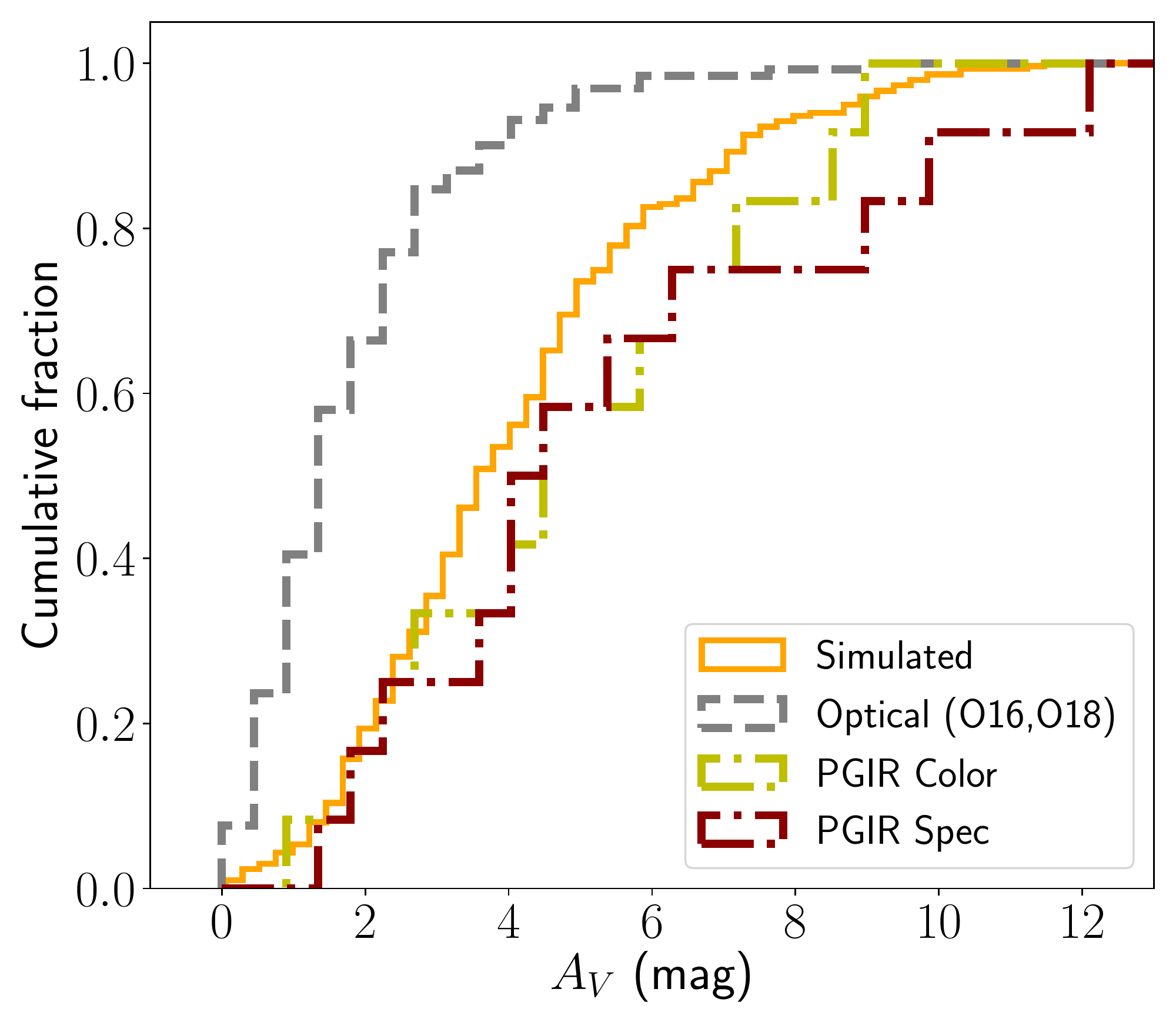}
    \caption{Comparison of the extinction ($A_V$) distribution of the PGIR sample (both photometric color and spectroscopic absorption estimates) of novae to that in previous optically selected samples (total of 131 objects) from \citealt{Ozdonmez2016} (O16) and \citealt{Ozdonmez2018} (018). We only include optically selected novae in the footprint accessible to PGIR for comparison. In cases where reliable spectroscopic estimates are not available, we use the color-based extinction estimates in the spectroscopic distribution curve (see Table \ref{tab:nova_list}). Also shown is the expected extinction distribution derived assuming that novae trace the stellar mass distribution in the Galaxy and using 3D dust distribution models from \citet{Green2019}.}
    \label{fig:ext_dist}
\end{figure}

We compare the derived extinction distribution of this sample to previous optically selected objects. We refer to the work of \citet{Ozdonmez2016} and \citet{Ozdonmez2018}, who accumulated a complete sample of extinction estimates (see Appendix of \citealt{Ozdonmez2016} for details of individual objects) of ($\approx 180$) optically discovered novae with photometric or spectroscopic measurements until 2018. In Figure \ref{fig:galdist}, we show the sky distribution of the optically discovered novae compared to the PGIR nova sample, colored by the inferred extinction towards each nova. Figure \ref{fig:galdist} also shows the estimated mass-weighted dust extinction along the line of sight (see Section \ref{sec:gal_model}), in particular, noting the complex dust structures that correlate with the inferred extinction towards the novae.

Unlike the PGIR sample that is concentrated towards heavily extincted regions near the Galactic plane, Figure \ref{fig:galdist} shows that previous optically discovered novae appear to preferentially populate higher galactic latitudes while also exhibiting lower extinction. In order to quantify this striking bias, we compare the cumulative distribution of the extinction ($A_V$) for previous optically selected novae to the PGIR sample in Figure \ref{fig:ext_dist}. As shown, the distribution for the PGIR sample is distinctively skewed towards larger extinction values compared to the optical sample. Performing a Kolmogorov-Smirnov (KS) test as well as a Anderson-Darling (AD) test between the optical and PGIR NIR sample (both photometric and spectroscopic estimates), we find that the null-hypothesis probability of being drawn from the same underlying population is $< 0.01$\%. This provides strong evidence that the PGIR sample uniquely probes a population of highly reddened novae that have been largely missed in previous optical searches. 

\subsection{Simulated estimate for the Galaxy}
\label{sec:gal_model}

We now compare the extinction distribution in Figure \ref{fig:ext_dist} to realistic models for the Milky Way. We create a simulated population of 1000 novae assuming a simple model where the nova rate traces the Galactic stellar mass distribution. We use recent estimates of the stellar mass distribution from \citet{Cautun2020} calibrated to Gaia DR2, that consists of three distinct components -- a bulge, a thin disk and a thick disk \citep{Bissantz2002, Juric2008, McMillan2017}.  We use the derived parameters in Table 1 and 2 of \citet{Cautun2020} for a contracted halo model. We then utilize the mass distribution together with the Galactic dust distribution using the \texttt{Bayestar2019} model from \citet{Green2019} to estimate a mass-weighted extinction sky map from Earth in Figure \ref{fig:galdist}. Assuming novae roughly trace the Galactic stellar mass distribution, the mass weighted extinction map indicates the typical expected extinction towards novae along each line of sight.

For each nova in the simulated population, we also estimate the extinction towards the nova and create a simulated extinction distribution shown in Figure \ref{fig:ext_dist}. As shown, the distribution observed for the optically selected population is preferentially skewed to low extinction values while realistic estimates for the Galaxy suggest that more than 50\% of novae should be obscured by $A_V \gtrsim 4$\,mag. On the other hand, we find that the PGIR distribution exhibits the expected high extinction tail with $>50$\% of events obscured by $A_V\gtrsim 5$\,mag, and thus commensurate with the expected distribution. This result bolsters our suggestion that previous optical searches have likely missed a large population of highly reddened (and optically faint) novae. Alternatively, some reddened novae could be misidentified as dwarf novae in optical searches and not followed up with spectroscopy. This holds true in the case of the nova V6567\,Sgr which was independently reported as a dwarf nova candidate by the ASASSN survey\footnote{\url{http://www.astronomy.ohio-state.edu/asassn/transients.html}}, and in the case of V2000\,Aql which was independently reported by the MASTER survey \citep{Pogrosheva2020} as a faint Galactic optical transient and not followed up with spectroscopy before the bright NIR detection.

\section{The Galactic rate of novae}
\label{sec:rate}
\begin{figure*}[!htp]
    \centering
    \includegraphics[width=\textwidth]{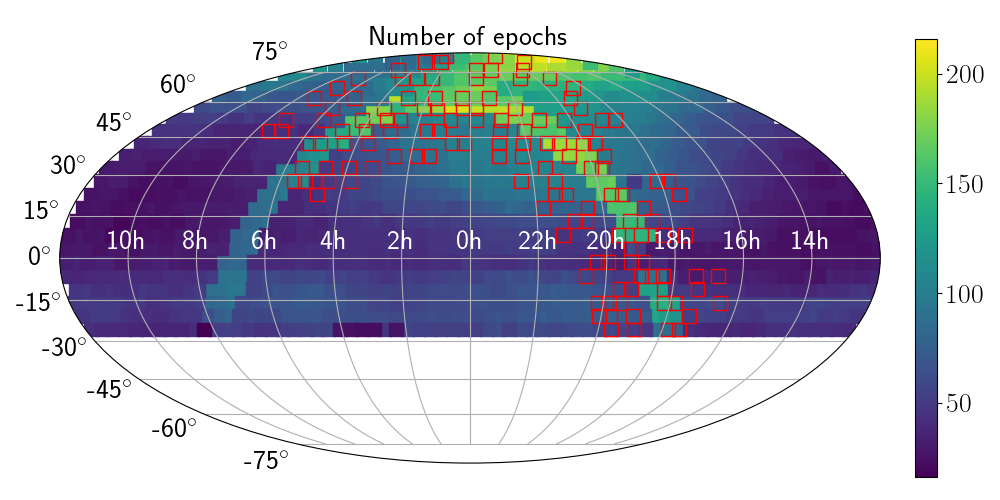}
    \caption{All-sky distribution of the number of epochs of observations from PGIR during the first 17 months of the survey. Fields at higher declination have larger number of epochs due to longer visibility over the year. Fields in the galactic plane also received a large number of visits due to a dedicated 1-day cadence observing program of the Galactic plane in 2020. The red squares show the sky positions of the fields where we carried out artificial source recovery tests.}
    \label{fig:epochs}
\end{figure*}

Having demonstrated that previous optical searches have likely missed a large fraction of novae, we turn to using the unique sensitivity of PGIR to dust obscured novae to constrain the Galactic rate. We proceed by using 11 out of the 12 novae that passed the selection criteria for our search, and estimate the Galactic rate assuming a simple model where novae trace the stellar mass distribution in the Galaxy. We discuss this assumption together with other additional effects in Section \ref{sec:discussion}.

\subsection{The pipeline detection efficiency}

Since the nova sample was identified using the candidates generated by the PGIR transient detection pipeline, we first quantify its detection efficiency. The overall detection efficiency has been demonstrated to be high, recovering $>90$\% of sources down to the $5\sigma$ limiting magnitude \citep{De2020a}. However, the efficiency is known to vary as a function of sky position due to varying amounts of source crowding and confusion given the large pixel scale of the detector. In order to quantify these biases, we selected 137 fields\footnote{The entire sky north of $\delta = -28.9^\circ$ is divided into 1329 PGIR fields.} distributed across galactic longitude in $0^\circ < l < 180^\circ$ and galactic latitudes in $-30 < b < 30^\circ$ to inject artifical sources into the images over a range of magnitudes. Since the detection efficiency is roughly constant at $|b| \geq 30^\circ$, we restrict our analysis to $|b| \leq 30^\circ$.

Figure \ref{fig:epochs} shows the sky positions of the fields where we injected artifical sources. For each selected field, we injected 200 artificial sources over a random selection of epochs, and distributed between the respective limiting and saturation magnitude. The artificial sources were injected in accordance with our nova identification criteria of requiring a detection at $>3$\,mags brighter than the nearest 2MASS source within 10\arcsec\ (if any). The images were then processed through the PGIR subtraction pipeline, and the detection efficiency (fraction of sources recovered) was quantified as a function of the relative brightness of the injected source with respect to the $5\sigma$ limiting magnitude of the image. Figure \ref{fig:recovery} shows the resulting recovery efficiency for a low and high Galactic latitude field. While the recovery efficiency is consistently high ($> 95$\,\%) at high latitudes, it is lower ($\approx 60 - 80$\%) at lower Galactic latitudes due to confusion \citep{De2020a}.

\subsection{Survey pointing simulations}

\begin{figure*}
    \centering
    \includegraphics[width=0.49\textwidth]{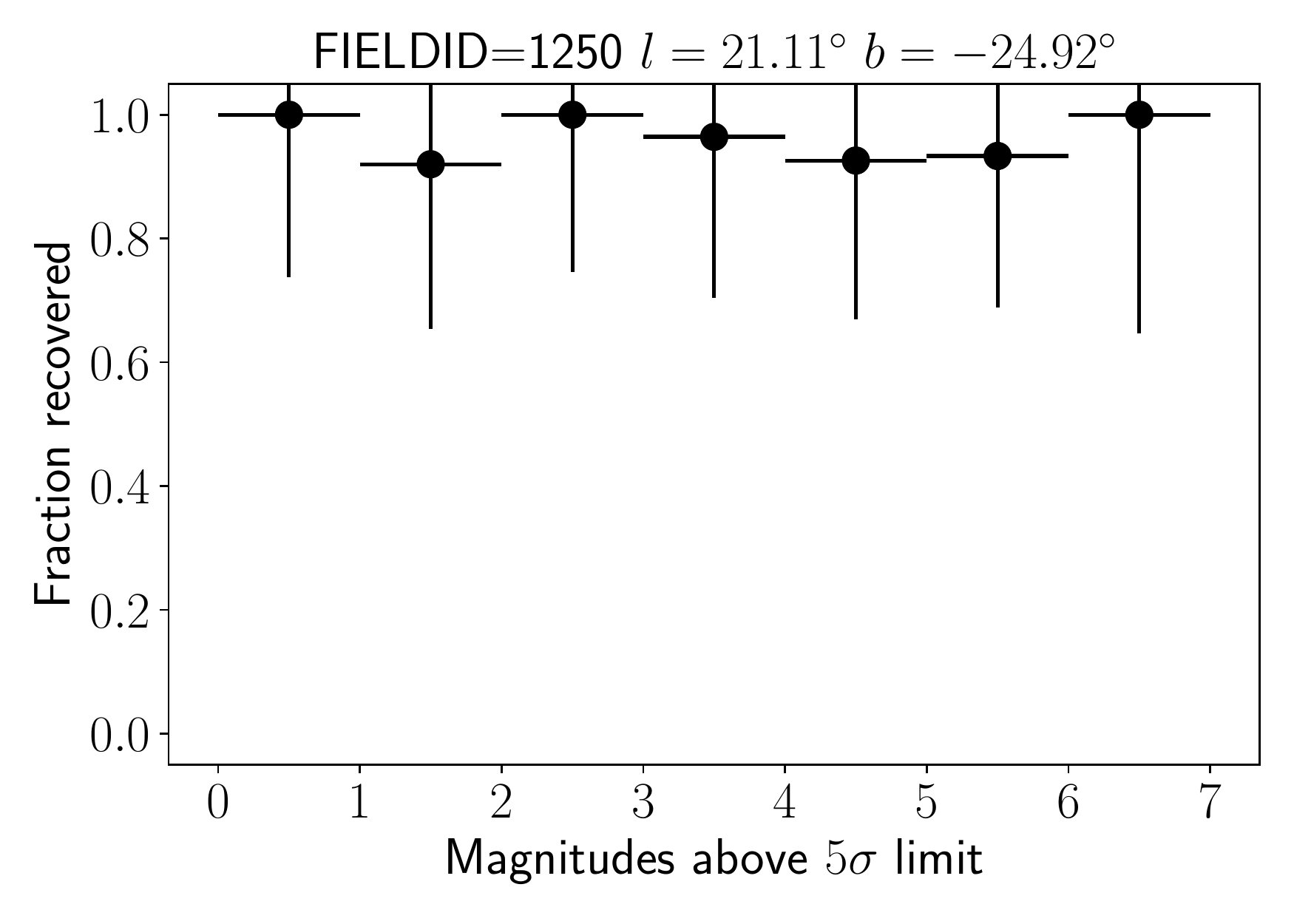}
    \includegraphics[width=0.49\textwidth]{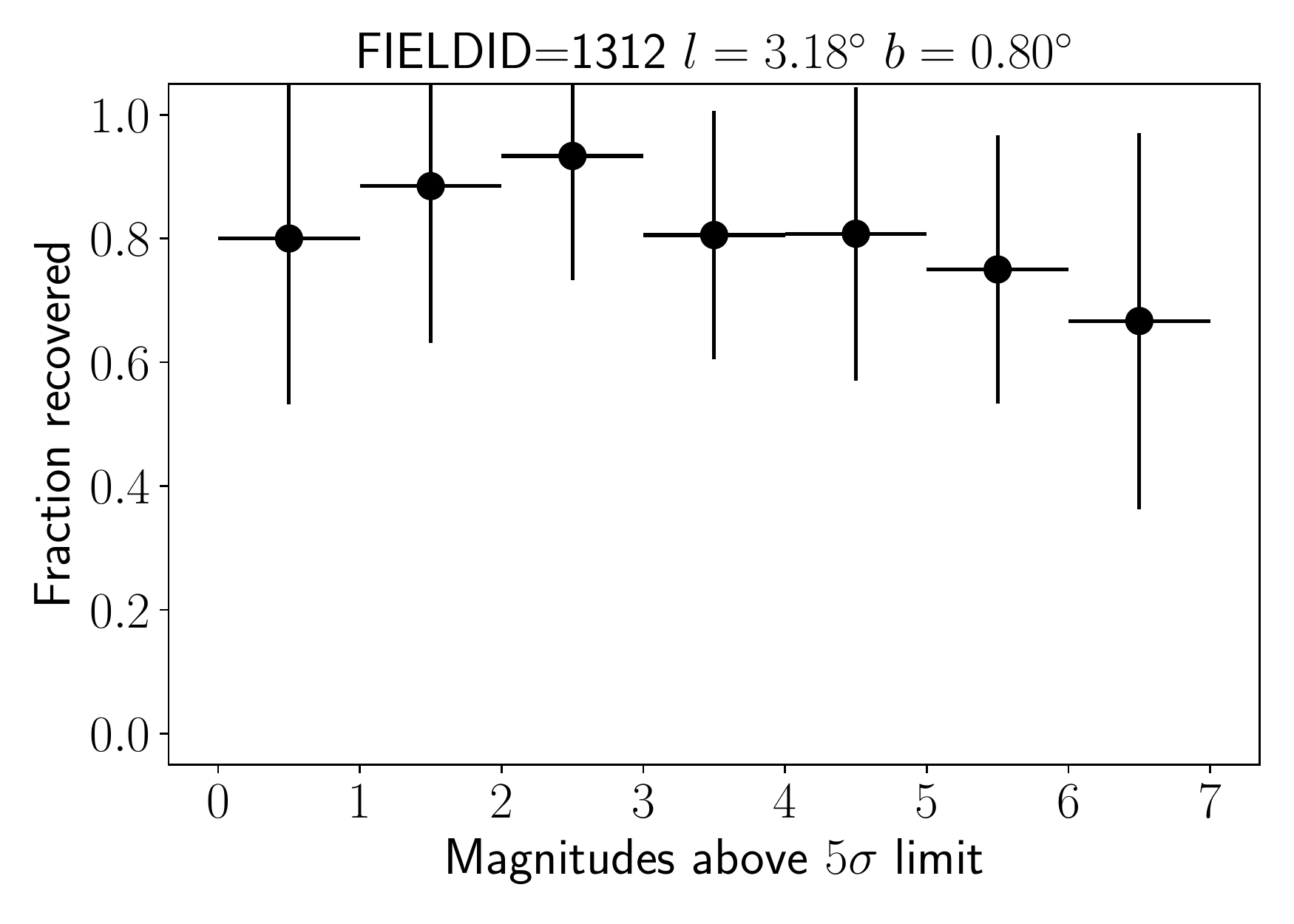}
    \caption{The detection efficiency of fake transient sources injected into the PGIR science images for two fields -- at high galactic latitude (left) and low galactic latitude (right). The panel titles show the PGIR field ID, galactic longitude ($l$) and latitude ($b$). We quantify the recovered fraction as a function of the difference between the fake magnitude and estimated limiting magnitude for the image (as a proxy for the expected signal-to-noise of the source) -- such that 0\,mag difference corresponds to sources exactly at the limiting magnitude, while larger values correspond to brighter transient sources. For high galactic latitudes (left), the recovery efficiency is roughly uniform at $\approx 95-100$\% for all sources brighter than the $5\sigma$ limiting magnitude. The recovery efficiency is reduced (typically $\approx 85$\%) at low latitudes due to confusion noise; the efficiency drops to $< 80$\% near the $5\sigma$ limit due to photon noise from nearby sources and at the bright end due to saturation induced by the bright stellar background.}
    \label{fig:recovery}
\end{figure*}

We simulate the recovery efficiency of novae in PGIR data utilizing the actual observing schedule together with the derived recovery efficiency of the PGIR subtraction pipeline. As in \citet{Shafter2017}, the peak magnitudes of the novae are drawn from the observed luminosity function of novae in M31 where the distance is very well constrained -- a normal distribution with a mean peak luminosity of $M = -7.2$\,mag and a standard deviation of $\sigma_M = 0.8$\,mag. For each nova with a peak luminosity, we estimate the rate of decline of the nova light curve using the Maximum Magnitude Rate of Decline (MMRD; \citealt{Zwicky1936, Mclaughlin1945}) relationship between the nova luminosity and speed from \citet{Ozdonmez2018}. We discuss possible deviations from the assumption of a universal luminosity function and MMRD in Section \ref{sec:discussion}.

Figure \ref{fig:epochs} shows the number of epochs of observations of PGIR as a function of sky position for the duration of the survey considered here. We simulate the actual survey pointing schedule of PGIR to estimate if a simulated nova would have been detectable in our data given its light curve shape and extinction along the line of sight.  A nova is detected in a simulated survey epoch if its brightness and sky location satisfy the same criteria we used to identify novae in the PGIR stream (Section \ref{sec:candidates}). For sources that pass this criteria, we randomly assign a detection or non-detection weighted by the derived detection efficiency (as a function of the brightness of the nova relative to the image limiting magnitude; Figure \ref{fig:recovery}) for the spatially nearest field. As in the real candidate identification criteria, we require at least three epochs of detections with SNR$> 8$ to classify a nova as recoverable in the PGIR survey.

\subsection{Monte-Carlo rate estimate}

\begin{figure*}
    \centering
    \includegraphics[width=0.49\textwidth]{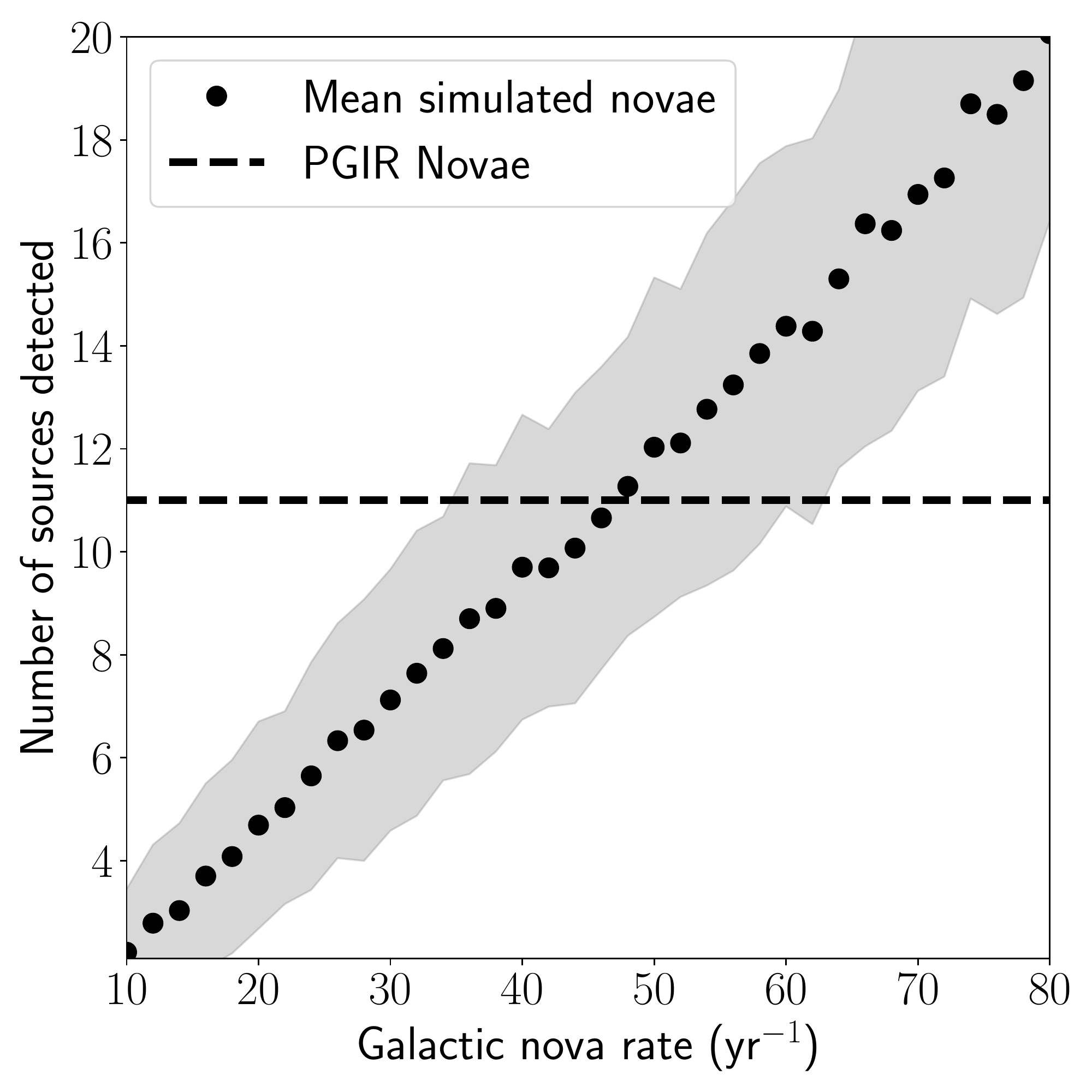}
    \includegraphics[width=0.49\textwidth]{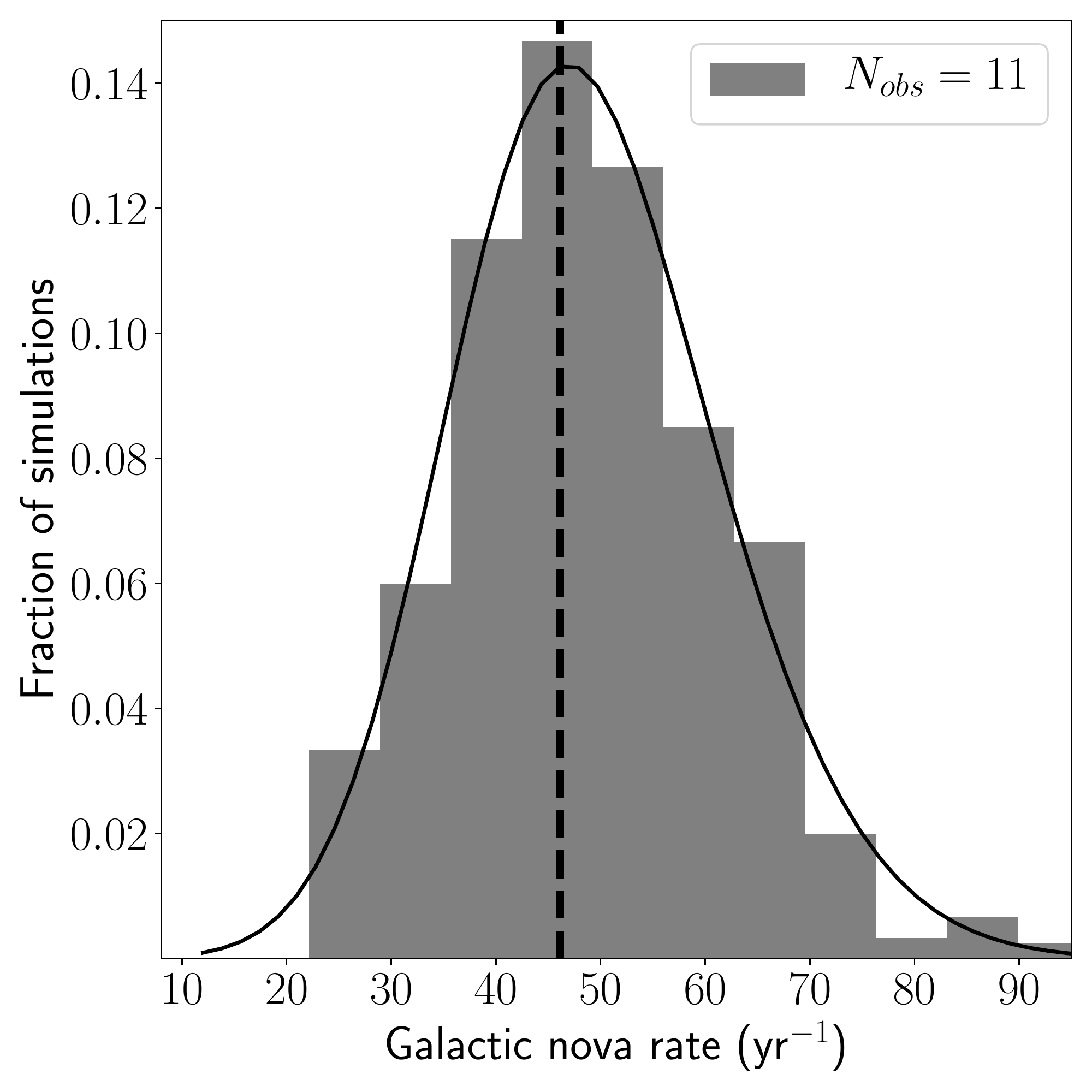}
    \caption{(Left) Simulated number of novae detectable in 17 months of the PGIR survey, as a function of the Galactic nova rate. Black circles show the mean number of novae detected for an input rate while the shaded region corresponds to the measured standard deviation. The black dashed line show the actual number of detected novae in PGIR. (Right) Distribution of the fraction of simulations that reproduce the observed number of PGIR novae as a function of the input Galactic rate. The solid black line shows the best-fit skewed Gaussian distribution and the black dashed line shows the best-fit rate. }
    \label{fig:rate_mc}
\end{figure*}

Using the framework for simulating the recovery of novae, we carry out a Monte Carlo simulation of the number of novae detectable in 17 months of the PGIR survey as a function of the input Galactic rate. For each input rate $r_0$, we create a population of $N$ novae  with locations weighted by the Galactic stellar mass distribution (Section \ref{sec:gal_model}), with $N$ randomly drawn from a Poisson distribution with a mean of $\lambda = r_0 * t_s$, where $t_s$ is the survey simulation duration in years:
\begin{equation}
P(N, \lambda) = \frac{\lambda^N}{N!} e^{-\lambda}
\end{equation}
Estimating the sky position of each simulated nova as viewed from Earth, we repeat the survey simulation for 100 iterations of each input Galactic nova rate ranging from $2$\,yr$^{-1}$ to $100$\,yr$^{-1}$. For each input nova rate, we record the mean number number of novae detected in the PGIR simulations as well as its standard deviation to estimate the uncertainty. Figure \ref{fig:rate_mc} shows the number of novae recovered in the first 17 months of the PGIR survey simulation as a function of the input Galactic rate. Similar to the rate estimation procedure discussed in \citet{De2020b}, we estimate the best-fit rate and its confidence interval by creating a distribution of the fraction of simulations that produce the observed number of novae ($= 11$) as a function of the global rate. We fit a skewed normal function to this distribution to estimate a Galactic rate (with 68\% confidence intervals) of 
\begin{equation}
    r_{0,u} = 46.0^{+12.5}_{-12.4}\,{\rm yr}^{-1}
\end{equation}
where $r_{0,u}$ is the Galactic rate assuming that the specific nova rate and luminosity function is uniform for the bulge and disk populations. We discuss possible variations in our assumptions and how they affect our rate estimates in Section \ref{sec:discussion}.

\section{Discussion}
\label{sec:discussion}

Using a sample of 12 spectroscopically confirmed novae detected in the PGIR survey, we have thus far demonstrated that the near-infrared sensitivity of PGIR has enabled the identification of a large population of highly obscured novae that have been systematically missed in previous optical searches. Unlike all previous estimates of the Galactic nova rate that suffer from poorly quantified completeness estimates, we derive an estimate of the Galactic nova rate combining i) a detailed quantification of the detection efficiency of the survey and ii) Monte Carlo simulations of the PGIR survey pointing schedule. In this section, we discuss the detection efficiency of the PGIR survey for novae in different parts of the galaxy, and revisit possible variations in the assumptions regarding the underlying population. 

\subsection{Detection efficiency for disk and bulge novae}

\begin{figure}[!ht]
    \centering
    \includegraphics[width=\columnwidth]{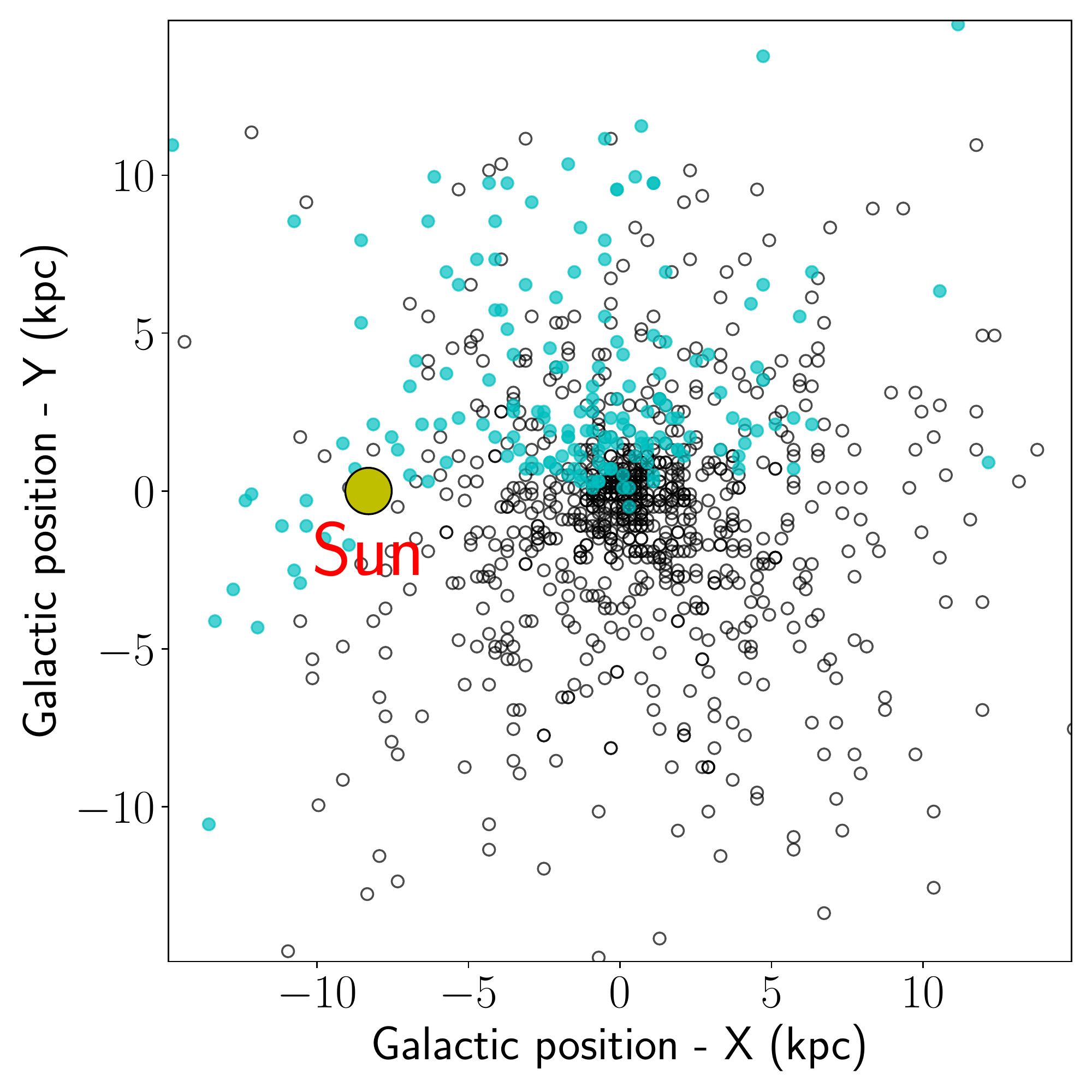}
    \caption{Nova detection capability in the PGIR survey as a function of galactic position. For a simulated population of 1000 novae, the cyan circles show eruptions that are detectable with the survey schedule and selection criteria while the black empty circles show the un-detected novae. Novae at $Y \lesssim 0$\,kpc are undetectable because they lie at Galactic longitude $l < 0^\circ$ that is largely invisible from Palomar. The position of the sun is shown as the yellow circle.}
    \label{fig:galpos}
\end{figure}

Here, we examine the detection efficiency of novae as a function of Galactic position, as imposed by our selection criteria in the PGIR sample. For a simulated population of $1000$ novae distributed by the Galactic stellar mass density, we show in Figure \ref{fig:galpos}, the Galactic x-y positions of the novae colored by their recovery in the survey. The majority of novae at $l < 0^\circ$ are not detectable by the PGIR survey due to their location in the southern hemisphere at $\delta < 30^\circ$. For $l > 0^\circ$, the sensitivity of PGIR through regions of high dust extinction allows the detection of novae in the central regions of the Galaxy as well as behind the Galactic bulge. Overall, our selection criteria with PGIR recover $\approx 17$\% of all novae injected into our simulation, and $\approx 36$\% of novae visible in the PGIR observing footprint. Using the same framework, we estimate that the PGIR recovery efficiency for disk novae was $\approx 18$\% of all eruptions, while the same for bulge novae was $\approx 13$\% owing to its location further in the southern hemisphere. Similarly, the recovery efficiency of disk and bulge novae in the footprint visible to PGIR is estimated to be $\approx 35$\% and $\approx 27$\% respectively.

Next, we compare the predictions of our assumed model for the relative contributions of disk and bulge novae with the observed statistics. Since we nominally assume the nova rate to scale with the stellar mass density, bulge novae constitute $\approx 20$\% (the bulge mass fraction in our assumed mass model from \citealt{Cautun2020}) of the eruptions. When combined with the recovery efficiencies for the respective populations, we expect $\approx 1-2$ bulge novae and $\approx 10-11$ disk novae during our survey duration. For comparison, we note that at least 7 out of the 11 ($\gtrsim 65$\%) novae in our rate sample can be certainly associated to the disk population based on their sky locations (see Figure \ref{fig:galdist}), while the rest are consistent with being bulge novae in terms of projected sky location.

As discussed in \citet{Hatano1997}, without accurate distance estimates, many novae apparently in the bulge could be foreground disk novae. Due to the long-lived light curves of most objects in our sample, we do not yet have measurements of most of their decline times ($t_3$, the time to decline by 3 mags by peak). However, for the well sampled light curve of apparent bulge nova V659\,Sct, we note that the distance estimated from the MMRD would be $\approx 4 - 5$\,kpc, and thus consistent with a disk population rather than the bulge. Similarly, the very slow nova V6567\,Sgr is likely to be relatively low luminosity and hence a foreground event. While we do not have extensive photometric coverage of V6593\,Sgr due to its proximity to the sun at the time of eruption, data from the AAVSO International Database\footnote{\url{www.aavso.org}} suggests a moderately fast nova with $t_3 \approx 35$\,days similar to V3731\,Oph ($t_3 \approx 40$\,days). As such, we suggest that these two moderately fast novae may have distances consistent ($\approx 8-9$\,kpc) with the bulge, making the number of bulge novae commensurate with our nominal model predictions.

\subsection{Luminosity function and light curve shape}

Our rate estimate was derived assuming a luminosity function represented by a normal distribution with peak absolute magnitude of $M = -7.2 \pm 0.8$\,mag, based on the luminosity function of novae in M31 \citep{Shafter2017}. However, several previous works have highlighted possible differences between the luminosity function of M31 and Milky Way novae, as well as differences between disk and bulge novae. \citet{Shafter2009} and \citet{Ozdonmez2018} have suggested that Milky Way novae are more luminous than M31 novae with an absolute magnitude distribution of $M_V = -7.9 \pm 0.8$\,mag, although \citet{Shafter2017} suggest that this conclusion may be biased by bright Galactic disk novae that are easier to find. In order to quantify the effect of a possibly brighter population of Milky Way novae, we carried out our Monte Carlo simulations assuming the suggested Galactic luminosity function and find a marginally higher resulting nova rate of $51.5^{+15.7}_{-15.5}$\,yr$^{-1}$. Although we expect brighter novae to be easier to detect in a simulated survey (producing a lower inferred rate for a fixed number of observed novae), the inferred rate is higher in these simulations since the faster evolution of bright novae are harder to recover in the survey.

Next, we discuss possible differences between the bulge and disk nova populations. \citet{Duerbeck1990}, \citet{DellaValle1992} and \citet{DellaValle1998} have shown evidence of likely different disk and bulge populations distinguished by their light curve speed and spectroscopic classification, wherein luminous ($M_V = -8 \pm 0.8$\,mag) and fast He/N novae preferentially appear in the disk population while slow and faint ($M_V = -7 \pm 0.8$\,mag) novae preferentially appear in the bulge population. The differences have been attributed to differing underlying stellar populations since more massive WDs in the disk are expected to produce faster, luminous outbursts \citep{Shara1981, Livio1992}. We quantified the effect of possible differing populations by carrying out our simulations with these two distinct populations of novae. The resulting rate estimate is marginally higher at $57.7^{+16.0}_{-15.9}$\,yr$^{-1}$, but still consistent with our estimate assuming a uniform nova population. 

Next, we discuss the validity of the assumed maximum absolute magnitude relation with decline time (MMRD). Specifically, multiple recent studies have questioned the validity of the MMRD utilizing high cadence observations of extragalactic novae where interstellar absorption is much less severe and uncertain. \citet{Kasliwal2011} presented evidence for a class of faint and fast novae in M31 that deviated from the MMRD relationship, and consistent with the predictions of \citet{Yaron2005}. Similar conclusions were reported from {\it Hubble Space Telescope} observations of M87 \citep{Shara2016, Shara2017} and later observations of M31 by the Palomar Transient Factory \citep{Cao2012}.

Using Gaia DR2 observations of old Galactic novae, \citet{Schaefer2018} have suggested that the MMRD relationship should not be used owing to its poor consistency, although \citet{Selvelli2019} do find evidence for the MMRD (see also \citealt{DellaValle2020}). Given the general uncertainty regarding the MMRD, we note that a population of faint and fast Galactic novae would further increase the inferred Galactic rate owing to the difficulty in detecting them in magnitude-limited samples. Although the true fraction of these novae has not been quantified yet, $\approx 25$\% of novae presented in the sample of \citet{Shara2016} were shown to be faint and fast, suggesting that our estimate of the Galactic nova rate may be underestimated by at least a similar fraction, subject to differences in the stellar populations.

\begin{figure*}
    \centering
    \includegraphics[width=\textwidth]{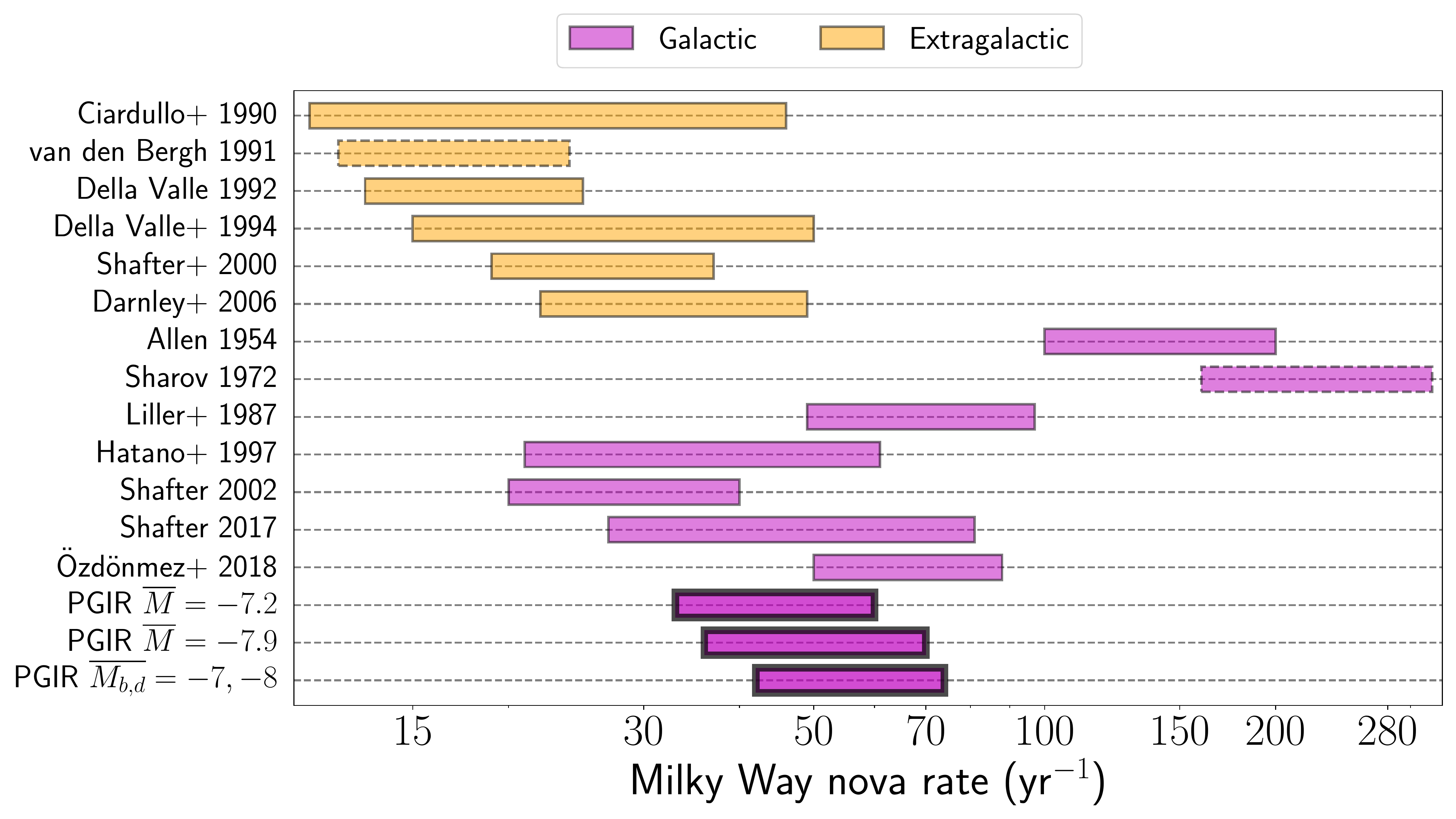}
    \caption{Comparison of previously published nova rate estimates for the Milky way to that estimated from the PGIR sample. Yellow bars indicate estimates from extrapolation of nova rates of nearby galaxies while magenta bars indicate those from samples of Galactic novae. Bars with dashed edges indicate values without published uncertainty estimates where we nominally assign a factor of two uncertainty. For the PGIR sample, we show the estimated range for three assumed luminosity functions: $M = -7.2 \pm 0.8$\,mag (as for M31 novae), $M = -7.9 \pm 0.8$\,mag (as suggested for Galactic novae) and in the case of two different luminosity functions in the bulge ($M_b = -7.0 \pm 0.8$\,mag) and disk ($M_d = -8 \pm 0.8$\,mag). }
    \label{fig:rate_compare}
\end{figure*}

\subsection{Comparison to previous estimates}

Previous estimates for the Galactic nova rate have primarily used the rate in the very local solar neighborhood to derive the local outburst rate density (although without a rigorous quantification of the completeness), followed by extrapolation to the entire galaxy (e.g. \citealt{Shafter2017, Ozdonmez2018}). Figure \ref{fig:rate_compare} summarizes our rate measurement in comparison to previous estimates. Our rate estimates are consistent with the work of \citet{Hatano1997} and the most recent work of \citet{Shafter2017}, although smaller than that estimated by \citet{Ozdonmez2018}. Compared to the Galactic bulge rate of $13.8 \pm 2.6$\,yr$^{-1}$ from \citet{Mroz2015}, the estimated bulge rate in our model would be $\approx 10 \pm 3$\,yr$^{-1}$, and marginally smaller than their estimate. Our estimates are inconsistent with the high rates ($> 100$\,yr$^{-1}$) estimated from the early work of \citet{Allen1954} and \citet{Sharov1972} using all-sky statistics of Galactic novae known at the time.

When compared to extra-galactic estimates, our derived rate is generally higher than those estimated in previous works in the range of $\approx 10 - 40$\,yr$^{-1}$. These underestimates likely arise from underestimation of the nova rate in external galaxies. For instance, \citet{Mroz2016} have shown that the nova rate in the Magellanic clouds is $\approx 2-3\times$ higher than estimated from the $K$-band luminosity using the OGLE survey. A similar high specific rate was inferred for M87 in the work of \citet{Shara2016}. While our estimate for the bulge rate is strikingly similar to that extrapolated from the M31 bulge by \citet{Darnley2006}, our estimate for the disk rate in the Milky Way ($\approx 38 \pm 12$\,yr$^{-1}$) is higher (but consistent within error bars) than their estimate of $\approx 20^{+14}_{-11}$\,yr$^{-1}$. The differences may be attributed to the different disk-to-bulge ratio in our model (constrained by recent Gaia DR2 data) compared to \citet{Darnley2006} together with differences in the stellar populations of the two galaxies. Indeed, the bulge population in M31 is known to be a more prolific contribution (per unit stellar light; \citealt{Ciardullo1987, Capaccioli1989, Shafter2001, Darnley2006}) to its nova rate as well as the stellar mass of the galaxy (disk-to-bulge luminosity ratio of $\approx 2$; \citealt{Shafter2002}).

On the other hand, bulge rate estimates from \citet{Mroz2015} together with previous works \citep{Hatano1997, Shafter2002, Shafter2017} have suggested that disk novae likely represent the majority of the nova outburst rate in the Milky Way. While our nominal model assumes a constant nova production rate per unit stellar mass and does appear to reproduce the observed number statistics, we are unable to further constrain the relative bulge-to-disk nova production rate due to low number statistics. Given the high extinction towards the bulge region, a larger sample of PGIR novae combined with upcoming datasets from NIR wide-field surveys in the southern hemisphere (with better visibility of the bulge) would be ideally suited to constrain the relative rates.

\section{Summary}
\label{sec:summary}

In this paper, we have presented a systematically selected sample of 12 spectroscopically confirmed novae detected in the first 17 months of the PGIR wide-field NIR survey. With $>50$\% of events obscured by $A_V \gtrsim 5$\,mag, this sample contains some of the most highly dust extinguished novae that have been spectroscopically confirmed in the literature. We use this sample to perform the first quantitative simulations of a time domain survey to directly constrain the Galactic nova rate. We summarize our findings below.

\begin{itemize}
    \item Comparing the extinction distribution of PGIR novae (derived from both photometric and spectroscopic tracers) to previously reported optical novae, we find the PGIR novae to be highly skewed towards large extinction values and inconsistent with the optical sample at $> 99.99$\% confidence.
    \item We create a simulated population of novae distributed by the stellar mass density in the Galaxy and estimate the extinction distribution towards the eruptions using recent 3D dust maps from \citet{Green2019}. We find the resulting extinction distribution to be commensurate with PGIR novae, suggesting that previous optical searches have likely missed or misidentified a large fraction of novae.
    \item We carry out detailed simulations of the PGIR pipeline detection efficiency together with the survey pointing schedule to estimate the Galactic nova rate. Using a simulated population of novae weighted by the stellar mass density and luminosity function (peak absolute magnitude $M = -7.2 \pm 0.8$\,mag), we estimate the Galactic nova rate to be $46.0^{+12.5}_{-12.4}$\,yr$^{-1}$.
    \item We further examined possible differences in our assumptions regarding the underlying luminosity function as well as differing bulge and disk populations. We derive marginally higher, but consistent rate estimates of $51.5^{+15.7}_{-15.5}$\,yr$^{-1}$ if Milky Way novae are characterized by a brighter luminosity function ($M = -7.9 \pm 0.8$\,mag). On the other hand, if disk and bulge novae have differing luminosity functions ($M = -8 \pm 0.8$\,mag and $M = -7 \pm 0.8$\,mag respectively), we derive an integrated rate of $57.7^{+16.0}_{-15.9}$\,yr$^{-1}$. The presence of faint and fast novae, as reported in some extra-galactic searches, would increase the inferred nova rate; our estimate thus serves as a lower limit.
    \item Our rate estimates are generally consistent with previous estimates extrapolated from novae only in the local solar neighborhood. However, our estimates are higher than those derived via extrapolations from nearby galaxies, which we attribute to previous underestimation of the specific nova rate in external galaxies as well as differences in assumptions regarding the relative contribution of bulge and disk novae.
\end{itemize}

Although all-sky optical surveys have consistently found only $5-10$ novae per year, our estimates are consistent with previous suggestions (which extrapolate the rate in the local solar neighborhood) of a Galactic rate of $\approx 50$\,yr$^{-1}$. Given the evidence for a population of highly reddened novae detected in PGIR that have been systematically missed in optical searches, our results suggest that the discrepancy in the observed rates arises due to dust obscuration preventing the discovery of most events in the optical bands. Wide and shallow NIR surveys are thus ideally suited to provide a complete census of the Galactic nova population. In particular, the lower effects of dust extinction in the NIR allow for easy discrimination between the abundance of faint (and nearby) dwarf novae and dust extinguished novae that are rarer but luminous in the NIR bands.

These results bode well for upcoming NIR surveys like the Dynamic REd All-sky Monitoring Survey (DREAMS; \citealt{Soon2020}), the Wide-field Infrared Transient Explorer (WINTER; \citealt{Simcoe2019}) and the Prime Focus Infrared Microlensing Experiment (PRIME). In particular, the finer pixel scale of these instruments as well as the southern locations of DREAMS and PRIME are suited to provide exquisite NIR statistics of novae in the highly dust extinguished and crowded southern Galactic bulge. Combining this population with the broader disk population from PGIR would provide accurate constraints on the long standing question of the relative bulge and disk nova rates in the Milky Way.

\section*{Acknowledgements}

We thank R. Williams for assistance with the identification of the NIR spectroscopic features. We thank M. Darnley, A. Shafter, M. Shara, R. D. Gehrz, L. Bildsten, E. S. Phinney and S. R. Kulkarni for valuable feedback on this work. We acknowledge with thanks the variable star observations from the AAVSO International Database contributed by observers worldwide and used in this research.

Palomar Gattini-IR (PGIR) is generously funded by Caltech, Australian National University, the Mt Cuba Foundation, the Heising Simons Foundation, the Binational Science Foundation. PGIR is a collaborative project among Caltech, Australian National University, University of New South Wales, Columbia University and the Weizmann Institute of Science. MMK acknowledges generous support from the David and Lucille Packard Foundation. MMK and EO acknowledge the US-Israel Bi-national Science Foundation Grant 2016227. MMK and JLS acknowledge the Heising-Simons foundation for support via a Scialog fellowship of the Research Corporation. MMK and AMM acknowledge the Mt Cuba foundation. J. Soon is supported by an Australian Government Research Training Program (RTP) Scholarship.

SED Machine is based upon work supported by the National Science Foundation under Grant No. 1106171. Some of the data presented here were obtained with the Visiting Astronomer facility at the Infrared Telescope Facility, which is operated by the University of Hawaii under contract 80HQTR19D0030 with the National Aeronautics and Space Administration. Some of the data presented herein were obtained at the W.M. Keck Observatory, which is operated as a scientific partnership among the California Institute of Technology, the University of California and the National Aeronautics and Space Administration. The Observatory was made possible by the generous financial support of the W.M. Keck Foundation. The authors wish to recognize and acknowledge the very significant cultural role and reverence that the summit of Mauna Kea has always had within the indigenous Hawaiian community. We are most fortunate to have the opportunity to conduct observations from this mountain.

This work was supported by the GROWTH (Global Relay of Observatories Watching Transients Happen) project funded by the National Science Foundation under PIRE Grant No 1545949. GROWTH is a collaborative project among the California Institute of Technology (USA), University of Maryland College Park (USA), University of Wisconsin Milwaukee (USA), Texas Tech University (USA), San Diego State University (USA), University of Washington (USA), Los Alamos National Laboratory (USA), Tokyo Institute of Technology (Japan), National Central University (Taiwan), Indian Institute of Astrophysics (India), Indian Institute of Technology Bombay (India), Weizmann Institute of Science (Israel), The Oskar Klein Centre at Stockholm University (Sweden), Humboldt University (Germany), Liverpool John Moores University (UK) and University of Sydney (Australia).  

\software{\texttt{astropy} \citep{Astropy2013}, \texttt{matplotlib} \citep{Hunter2007}, \texttt{scipy} \citep{Virtanen2019}, \texttt{pysedm} \citep{Rigault2019}, \texttt{pyraf-dbsp} \citep{Bellm2016b}, \texttt{spextool} \citep{Cushing2004}, \texttt{xtellcor} \citep{Vacca2003},
\texttt{lpipe} \citep{Perley2019}}

\facilities{PO: Gattin-IR, PO: 1.2m (ZTF), PO: 1.5m (SEDM), Hale (DBSP, TSpec), IRTF, Keck:I (LRIS), AAVSO.}

\appendix

\section{Summary of individual objects}
\label{sec:app_nova}

\begin{figure*}
    \centering
    \includegraphics[width=\textwidth]{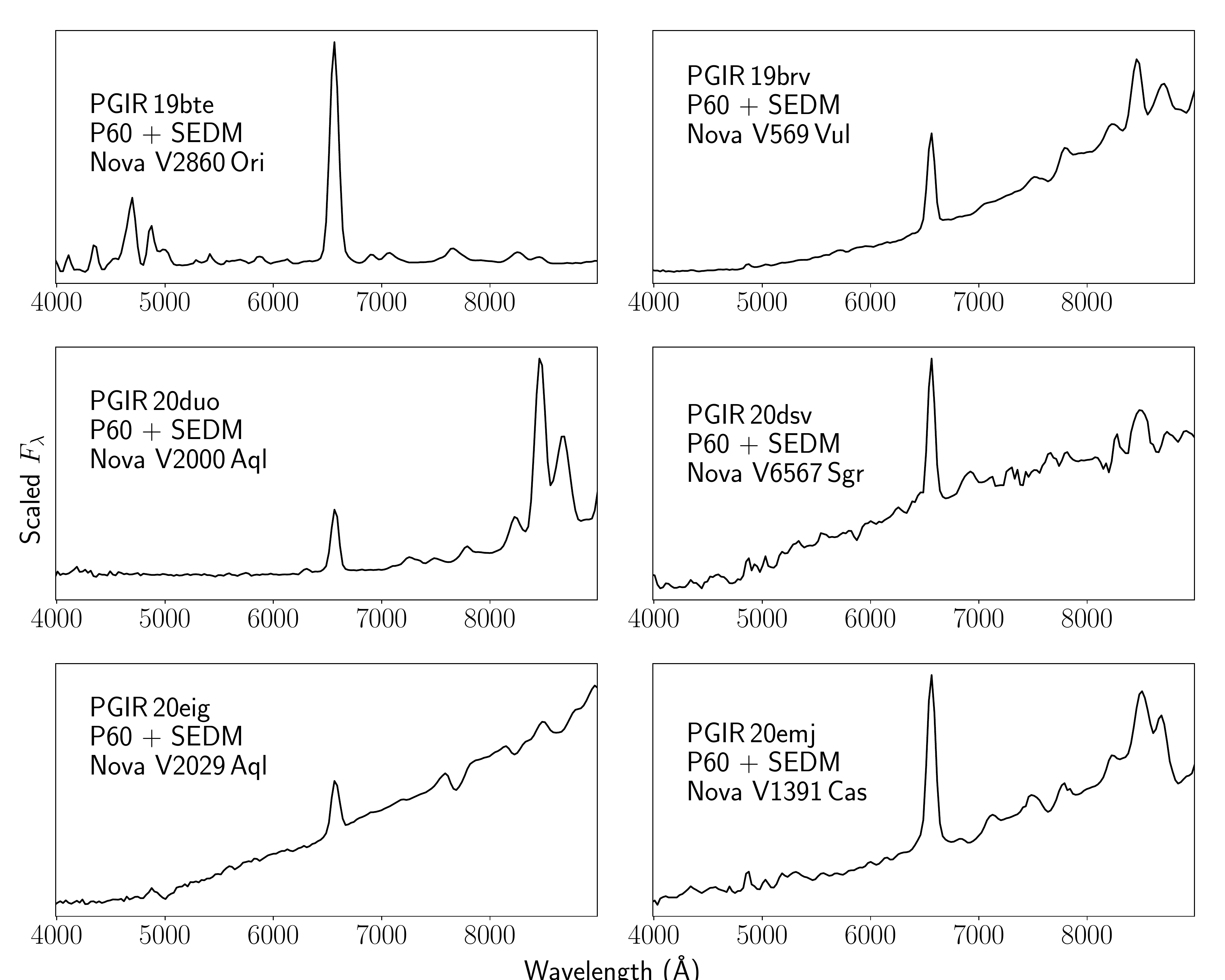}
    \caption{Rapid ultra-low resolution confirmation spectra of PGIR novae obtained with the SED Machine Spectrograph. Each spectrum (denoted with the nova name) shows strong H$\alpha$ emission together with multiple emission features of O I, confirming the nova classifications. }
    \label{fig:novae}
\end{figure*}

\subsection{Nova V3731 Oph}

The transient PGIR\,20ekz was recovered in an archival search for large amplitude transients (using the criteria described in Section \ref{sec:candidates}) in early survey data from July 2019 \citep{DeA2020d}. The source was also detected as a fast reddened transient ZTF\,19abgnhzj in ZTF public data, but was missed and not followed up during the eruption. We obtained a late-time optical spectrum of the source using Keck-I + LRIS on 2020-09-15 ($\approx 420$\,days from peak) which confirmed its classification as a highly reddened nova in the nebular phase. Due to the large diversity in nebular phase features of novae, we are unable to confidently constrain the spectroscopic type; however, taking the detection of He I, [N II], [Ne III] 3869/3968, [O II] 7325 and [S III] 9069 (Figure \ref{fig:spec_1}; similar to the very fast He/N nova V838\,Her in \citealt{Williams1994}) together with the fast evolving light curve, we tentatively suggest a He/N classification. The limited photometric coverage exhibits a smooth decline, and is consistent with a S-type photometric class. Due to the absence of continuum emission in the late-phase spectrum, we are unable to detect any spectroscopic features to estimate the reddening; we thus use the color evolution of the nova for our estimate. 

\subsection{Nova V2860 Ori}

The nova V2860\,Ori was first discovered by Shigehisa Fujikawa, Kan'onji, Kagawa, Japan as PNV J06095740+1212255 on UT 2019-08-07, and classified by \citet{Aydi2019a} as fast classical nova. The transient was recovered in PGIR data as PGIR\,19bte with our selection criteria, and confirmed with rapid SEDM spectroscopy (Figure \ref{fig:novae}). With follow-up optical and NIR spectroscopy on the P200, we confirm its classification as a He/N nova based on the detection of He\,I and N\,II features with broad, flat-topped emission profiles (Figure \ref{fig:spec_1}). Combining the photometric coverage from PGIR and ZTF, we find a smooth decline in the nova light curve followed by a dramatic dip from dust formation at $\approx 150$\,days after outburst, suggesting a D-type photometric class. Using the detect K I $\lambda7699$ absorption feature in the DBSP spectrum, we estimate an extinction of $E(B-V) \approx 0.55$\,mag. 

\subsection{Nova V569 Vul}

The nova V569\,Vul was first reported by the Gaia survey \citep{Hodgkin2019a} on UT 2019-08-24 and spectroscopically confirmed as a highly reddened classical nova \citep{Aydi2019b, Zielinski2019}. Due to the highly reddened nature of the source, the nova was recovered as a very bright J-band transient (PGIR\,19bgv) in the PGIR data, and brighter than the nominal non-linearity limit ($\approx 8.5$\,mag) for the first two epochs. We obtained spectroscopic follow-up of the source using P200 and classify the source as a fast He/N nova based on the detection of He\,I and N\,II features with flat-topped emission profiles (Figure \ref{fig:spec_1}). The extremely red color of the outburst (Figure \ref{fig:photometry}) suggests a large extinction ($A_V \approx 10$\,mag), and is consistent with the high reddening we measure from the optical spectrum using the K I $\lambda7699$ absorption line. The light curve exhibits a smooth decline from outburst peak, and consistent with a S-type photometric class.

\subsection{Nova V2891 Cyg}

The nova V2891\,Cyg was first discovered by the PGIR survey as the transient PGIR\,19brv \citep{DeA2019} on UT 2019-09-17 and confirmed with rapid spectroscopic follow-up on the Palomar 60-inch telescope (Figure \ref{fig:novae}). The nova exhibited a smooth rise to peak followed by at least a $\approx 100$\,day bumpy plateau, suggesting a F-type photometric class. With higher resolution spectroscopic follow-up on the P200, we detect a highly reddened spectrum with Fe II features and P-Cygni line profiles (Figure \ref{fig:spec_1}), suggesting a Fe II-type spectroscopic classification. Both the highly reddened colors as well as the K I $\lambda7699$ spectroscopic feature suggest a very high reddening -- we estimate a reddening of $A_V \approx 7.5$\,mag from the colors and higher estimate of $A_V \approx 12$\,mag from the K I feature. We caution that the K I feature has not been calibrated at such high extinctions and likely saturates in this regime.

\subsection{Nova V3890 Sgr}

The nova V3890\,Sgr is a known recurrent nova \citep{Schaefer2010} with the 2019 eruption \citep{Strader2019} detected in PGIR data (as PGIR\,19fai). Owing to saturation of the bright eruption in survey data, it does not pass our selection criteria (Section \ref{sec:candidates}) and not included for our rate estimates. We obtained optical and NIR spectroscopy of the transient with P200, and detect broad H$\alpha$ and higher Balmer emission from the nova, together with clear absorption features from the stellar companion (Figure \ref{fig:spec_1}). With very high resolution spectroscopy of the 2019 eruption, \citet{Munari2019} estimated a line-of-sight extinction of $E(B-V) \approx 0.62$\,mag, which we use in our work.

\subsection{Nova V659 Sct}

The nova V659\,Sct was discovered by the ASASSN survey \citep{Stanek2019} on 2019-10-29, and confirmed with optical spectroscopy shortly thereafter \citep{Williams2019}. The nova was detected around the time of peak eruption in PGIR data but was brighter than the nominal non-linearity limit until it went into solar conjunction. The nova subsequently passed our selection criteria as PGIR\,20dcl and followed up with optical and NIR spectroscopy with SSO and IRTF. Based on the detection of Fe II features as well as N II / He I features in the optical spectra (Figure \ref{fig:spec_1}), we classify the source as a hybrid nova. The nova exhibited a smooth photometric decline until $> 200$\,days after eruption, followed by a phase of erratic bumps on top of the smooth decline. We thus obtain a photometric classification of a J-type nova based on its light curve. Using the detection of K I $\lambda 7699$ absorption in the optical spectrum, we estimate a reddening of $A_V \approx 4.0$\,mag, similar to that estimated from the photometric colors.

\subsection{Nova V2000 Aql}

The nova V2000\,Aql (= PGIR\,20duo) was first detected in the PGIR data on UT 2020-05-12, and identified as a nova candidate on its second detection on 2020-06-18. We obtained rapid spectroscopic follow-up of the source using SEDM (Figure \ref{fig:novae}) to confirm a nova classification \citep{DeA2020b}. The source had been previously reported as a faint hostless transient by the MASTER survey \citep{Pogrosheva2020} but not followed up until the PGIR identification. Combining PGIR data with public ZTF data, we find the nova to be highly reddened with $g - J \approx 10$\,mag. While the first observation of the nova was after a seasonal gap in PGIR data, the ZTF data constrain the eruption to have occurred $\approx 40$\,days before the first PGIR detection. With optical and NIR spectroscopic follow-up, we find evidence of a highly reddened nova of the Fe-II type based on the clear detection of C I lines (Figure \ref{fig:spec_1}) in the NIR spectra \citep{Banerjee2012}. We classify the photometric behavior of the nova as a S-type object based on the smooth decline seen in the NIR and optical data. Due to the highly reddened nature of the source, we do not detect any clear extinction features in the optical spectra and hence constrain the reddening using only the photometric colors to be $A_V \approx 9.5$\,mag. 

\subsection{Nova V6567 Sgr}

The nova V6567\,Sgr (= PGIR\,20dsv) was discovered by PGIR on 2020-06-01 \citep{DeA2020a} and confirmed with rapid SEDM spectroscopy (Figure \ref{fig:novae}). Using optical spectroscopy shortly after eruption, we classify the nova as Fe-II type based on the detection of multiple Fe features in the optical spectra and P-Cygni line profiles (Figure \ref{fig:spec_1}). The nova exhibits a slow decline superimposed with oscillations with a period of $\approx 30$\,days leading to its photometric classification as a O-type nova. Using the clear detection of the DIB $\lambda5780$ absorption in its optical spectrum, we estimate a spectroscopic reddening of $\approx 4.5$\,mag, consistent with that estimated from its photometric colors.

\subsection{Nova V2029 Aql}

The nova V2029\,Aql was discovered in PGIR data on UT 2020-07-13 \citep{DeA2020c}, and confirmed with SEDM spectroscopy on 2020-08-02 (Figure \ref{fig:novae}). The nova (PGIR\,20eig) exhibited a peculiar light curve with a relatively smooth rise for $\approx 50$\,days since eruption followed by a smooth drop in brightness. The smooth decline for $\approx 30$\,days was interrupted by rapid oscillations in its light curve over $\approx 1$\,day timescales \citep{Babul2020}. Based on the cusp-like peak in the light curve, we suggest its photometric type as a C-class nova. We obtained optical and NIR spectroscopy of the nova, which suggest a Fe-II type classification based on the very low ($\approx 400$\,km\,s$^{-1}$) P-Cygni line velocities and clearly detected Fe II features (Figure \ref{fig:spec_1}). Using the DIB $\lambda5780$ feature in the optical spectra, we estimate a reddening of $A_V \approx 6.0$\,mag, consistent with the photometric color estimate.

\subsection{Nova V1391 Cas}

The nova V1391\,Cas was discovered on 2020-07-27 UT by S. Korotkiy as part of the NMW survey, and classified as a classical nova with spectroscopic follow-up \citep{SokolovskyA2020}. The nova was detected in PGIR data on 2020-08-15 as PGIR\,20emj, and confirmed with a SEDM spectrum (Figure \ref{fig:novae}). With higher resolution optical and NIR spectroscopic observations, we obtain a spectroscopic classification as a Fe II nova based on the detection of Fe II features and P-Cygni line profiles in the optical spectra (Figure \ref{fig:spec_1}). The nova light curve exhibits a smooth decline for $\approx 120$\,days after eruption before a dramatic drop in brightness accompanied by reddening of its colors. We attribute this transition to dust formation, classifying this source a D-type nova. The onset of dust formation is detected on a timescale similar to that seen for dust formation in the recent bright D-type nova V5668 Sgr \citep{Gehrz2018},which also show a similar pre-dust-formation light curve. Using the detection of the DIB $\lambda5780$ absorption feature, we estimate a spectroscopic reddening of $A_V \approx 4.5$\,mag consistent with the estimate derived from photometric colors.

\subsection{Nova V6593 Sgr}

The nova V6593\,Sgr was detected in the PGIR data on UT 2020-10-02 (as PGIR\,20evr) and independently reported as a nova candidate by the BrATS survey on UT 2020-10-03 \citep{Jacques2020}. The bright NIR transient was subsequently confirmed as a Fe-II type classical nova with optical spectroscopy \citep{AydiA2020}. Owing to the short visibility of the nova before solar conjunction, PGIR only covered the rise of the nova light curve to peak, while we obtained only one higher resolution NIR spectrum of the source with P200. The NIR spectrum shows strong features of C\,I (Figure \ref{fig:spec_1}), consistent with the classification as a Fe-II type nova \citep{Banerjee2012}. We used the photometric color of the source obtained with an image from the P60 SED Machine camera to estimate an extinction of $A_V \approx 4$\,mag.

\begin{figure*}[!ht]
    \centering
    \includegraphics[width=\textwidth]{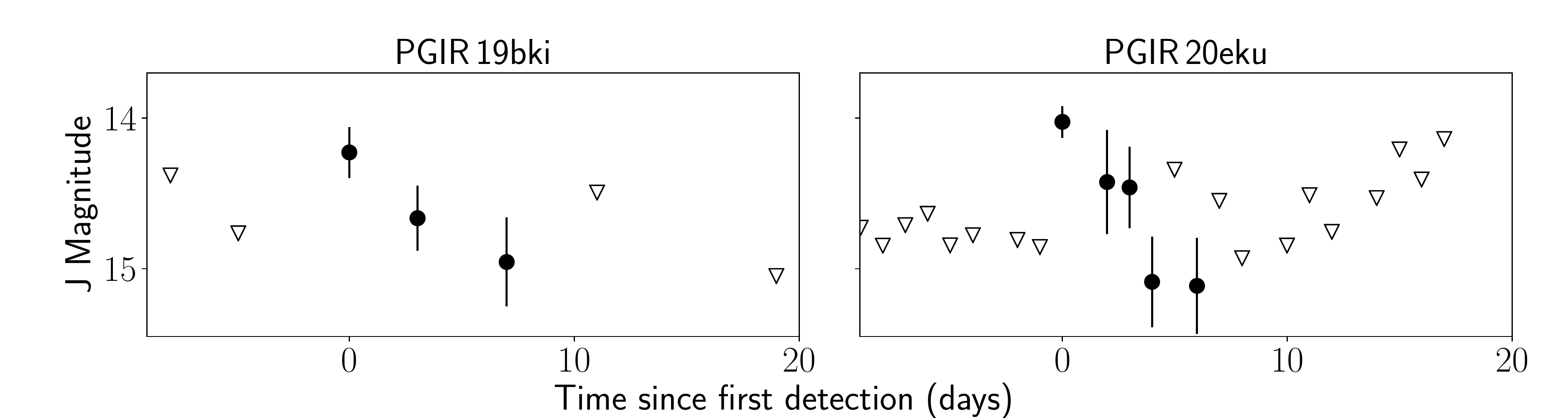}
    \includegraphics[width=\textwidth]{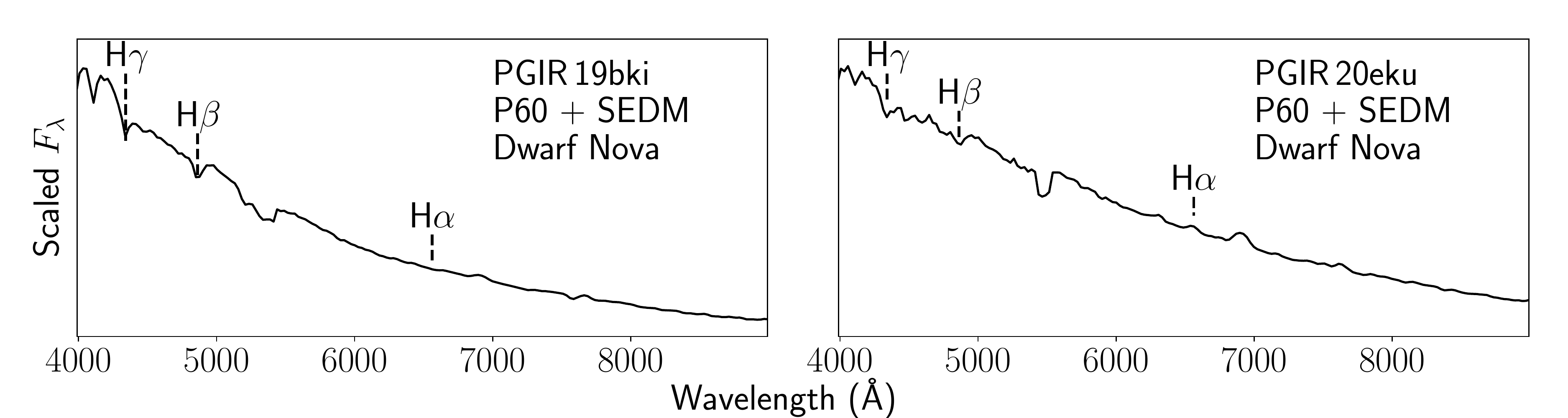}
    \caption{Light curves and spectra of dwarf novae detected and followed up as part of the PGIR nova search. The top panel shows the light curves of PGIR\,19bki and PGIR\,20eky, while the lower panel shows follow-up confirmation spectra from the SED Machine. The spectra show clear signs of weak H$\alpha$ features in emission and higher Balmer features in absorption, confirming their classification as dwarf nova outbursts.}
    \label{fig:dne}
\end{figure*}

\subsection{Nova V1112 Per}

Nova V1112\,Per was discovered on UT 2020-11-25 as TCP J04291884+4354232 by Seiji Ueda (Kushiro, Hokkaido, Japan) and confirmed with optical spectroscopy by \citet{Munari2020a}. The nova was detected as a bright NIR transient in PGIR data as PGIR\,20fbf, and followed up with optical and NIR spectroscopy on the P200. Based on the detection of multiple Fe II lines and P-Cygni line profiles in the optical spectrum (Figure \ref{fig:spec_1}), we classify the source as a Fe II nova. However, rapid early-time spectroscopic monitoring of the nova has revealed `reverse hybrid' (R-Hybrid) behavior \citep{Munari2020a, Borthakur2020} wherein the nova transitioned from a He/N appearance to Fe II appearance (seen at the epochs of our spectra) in a sense opposite to that seen in hybrid novae \citep{Williams1992}. Following a smooth decline from peak, the nova exhibited signatures of dust formation (hence classified as D-class) via rapid reddening of its optical-NIR color in the combined multi-color light curve. We estimate an extinction of $A_V \approx 2.5$\,mag using the J-band data from PGIR with optical photometry reported by \citet{Munari2020a}, and consistent with the spectroscopic estimate derived using the K I $\lambda7699$ absorption line in the optical spectrum.

\subsection{Dwarf nova PGIR 19bki}

The transient PGIR\,19bki was discovered by PGIR on UT 2019-09-12 at coordinates RA = 19:53:46.5, Dec = -07:48:38. The source was identified as a fast fading large amplitude transient (Figure \ref{fig:dne}) without any history of prior outbursts in other time domain surveys or in SIMBAD, suggesting a possible clasical nova outburst. Rapid spectroscopic follow-up on P60 + SEDM revealed a blue and largely featureless spectrum characterized by weak H$\alpha$ emission and higher order Balmer absorption features (Figure \ref{fig:dne}). We thus classify the source as a dwarf nova outburst.

\subsection{Dwarf nova PGIR 20eku}
 
The transient PGIR\,20eku was discovered by PGIR on UT 2020-08-04 at coordinates RA = 03:52:37.9, Dec = +47:51:05.9. The source was identified as a fast fading large amplitude transient (Figure \ref{fig:dne}) without any history of prior outbursts in other time domain surveys or in SIMBAD, suggesting a possible clasical nova outburst. Rapid spectroscopic follow-up on P60 + SEDM revealed a blue and largely featureless spectrum characterized by weak H$\alpha$ emission and higher order Balmer absorption features (Figure \ref{fig:dne}). We thus classify the source as a dwarf nova outburst.

\end{document}